\documentclass[journal,12pt,onecolumn,draftclsnofoot]{IEEEtran}
\IEEEoverridecommandlockouts

\usepackage{cite}
\usepackage{amsmath,amssymb,amsfonts}
\usepackage{algorithm}
\usepackage{algorithmic}
\usepackage{graphicx,psfrag}
\usepackage{graphicx}
\usepackage{textcomp}
\usepackage{xcolor}
\usepackage{amsthm}
\usepackage[justification=centering]{caption}
\usepackage{overpic}
\usepackage{multirow}
\usepackage{fancyhdr}
\usepackage{lmodern}

\usepackage{bm}
\usepackage{makecell}
\usepackage{setspace}
\usepackage{caption}
\title{RIS Aided Multi-Way Communication Systems based on SIC Decoding Techniques}
\author{Lingya Wang \footnote{Lingya Wang  was with the Department of Electrical and Computer Eng. McGill University, Montreal QC Canada. She is is now with China Telecom, P.R. China. Email: lingya.wang@mail.mcgill.ca}
 and
 Harry Leib \footnote{Harry Leib is with the Department of Electrical and Computer Eng., McGill University, Montreal, QC Canada, Email: harry.leib@mcgill.ca}	}

\begin{document}

\maketitle

\begin{abstract}
This work considers  reconfigurable intelligent surface (RIS) aided multi-way communication systems employing QPSK. The objectives of such system is to allow multiple users to exchange information, exploiting the full-duplex capabilities of RIS. The system under consideration comprises $K$ user terminals and  an RIS with $N$ reflective elements. Each user terminal is equipped with one transmit  and $M$ receive antennas. To mitigate inter-user interference caused by the full-duplex operation of the RIS, three successive interference cancellation (SIC) techniques are considered. We investigate three SIC  ordering methods: $1)$ descending channel variances, $2)$ increasing mean square error, $3)$  optimization approach which maximizes the combined channel gain within the system. The bit error rate (BER) performance of these three SIC methods are evaluated using extensive Monte-Carlo simulations  for different numbers of RIS elements, ranging from 4 to 64,  and a varying number of transmitting users, ranging from 1 to 5.  Additionally, an RIS phase shift optimization method  for RIS with  reflection elements of varying amplitudes is considered. In this case  the previous SIC optimization technique is adapted  to address  phase-dependent amplitudes in RIS reflection elements. Simulation results  indicate that in both ideal and varying amplitude scenarios, when the number of RIS elements is small, SIC with ordering  based on increasing mean square error achieves the best BER performance. However, as the number of RIS elements increases,  SIC ordering based on the optimization method achieves the best BER performance.
\end{abstract}


\begin{spacing}{1.7}

\section{Introduction}

\IEEEPARstart{I}{n} contemporary wireless communication environments, the demand for higher frequencies and peak data rates is on the rise. To address such challenging requirements, the concept of  reconfigurable intelligent surface (RIS) technology emerges as primary approach \cite{An2026}
  \cite{tut2026}. An RIS system  enables  incident EM waves to be reflected in a controllable way, generating a controllable wireless propagation environment \cite{Alomainy_2026} \cite{renzo2020}. A passive RIS system consists of a number of passives reflecting elements, each capable of independently altering the phase and/or amplitude of an  incident signal \cite{wu2021intelligent}. Consequently, RIS aided wireless communication systems can mitigate  signal attenuation or blockage caused by  obstacles such as buildings and trees\cite{zeng2022joint} resulting in improved  performance \cite{Khan2026}.

The multi-way communication paradigm is a generaliztion  of conventional two-way communication system\cite{van1977survey}\cite{eswaran2008achievable}, \cite{gunduz2012multiway}, which was proposed by Shannon in \cite{shannon1961two}. The objective of  multi-way communication is to achieve full-mutual data exchange among multiple users\cite{li2020intelligent}.  Multi-way communications techniques find applications in numerous systems such as  satellite communication\cite{ong2011capacity} and sensor network\cite{amarasuriya2013multi}. The use of RIS technology for multi-way communication seems as a promissing approach to enhance bandiwth efficiency of such systems  \cite{lwang2025} exploting the full duplex capabilities of such technology \cite{liu2026enabling} \cite{perveen2025}.

Current research progress on RIS includes the architecture realization of RIS\cite{di2020reconfigurable} at low cost, size, weight, and power (C-SWaP).  Recent implementations of RIS use either inexpensive antennas or metamaterials. For example, inexpensive antenna of which the inter-distance is of the order of half the wavelength can be used to realize large arrays\cite{di2020reconfigurable}\cite{li2019towards}. Major applications of RIS include waveform shaping for anomalous reflection/transmission synthesis \cite{pan2021full}, beamforming/focusing\cite{sugiura2021joint}\cite{yurduseven2020intelligent}, joint transmitter/RIS encoding\cite{karasik2020beyond} and single-RF multi-stream transmitter design\cite{jamali2020intelligent}\cite{tang2020wireless}\cite{pei2021ris}.  When using RIS, the wireless networks can be regarded as a smart radio environment \cite{ayoubi2025} which can adapt to various system conditions. Comprehensive introductions to signal models for RIS aided SISO, SIMO, and MIMO systems are provided in \cite{wu2021intelligent}. Furthermore, in \cite{zeng2022joint}, a RIS aided multi-antenna MU MIMO downlink transmission system with one BS equipped with $N_t$ transmit antennas, one RIS with $N$ reflecting elements, and $K$ users each with $N_r$ receive antenna is proposed. The RIS aided multi-user system in \cite{abeywickrama2020intelligent} considers a practical RIS phase shift model which captures the relationship between the reflection amplitudes and phase shifts.

When using the full duplex features of RIS technology in a multi-user system, inter-user interference creates a major challenge.To alleviate the effects of such iterference,  SIC techniques constitute a practical approach \cite{li2020intelligent} \cite{yue2023} \cite{kumar2025}. A significant factor in SIC decoding is the order of which users' data is detected. Several criteria exist for SIC decoding. For instance, over MIMO channels, ZF SIC and MMSE SIC detection  are widely utilized techniques \cite{choi2010optimal}. To maximize the minimum achievable rate of a multi-way system, sorting based on channel power coefficients is suggested in \cite{li2020intelligent}. A SIC decoding order based on RIS beamforming optimization is proposed in \cite{li2021joint}, which aims at minimizing BS transmit power. To maximize the system throughput, a low-complexity SIC decoding order optimization algorithm is proposed in \cite{zuo2020resource}. 

The models in \cite{li2021joint} and \cite{zuo2020resource} focus on downlink systems involving transmission from a BS to multiple receivers. Our work, considers a multi-way  communication model where transmission occurs among all users. Furthermore, in many papers related to RIS aided systems, the focus is principally on the performance characteristics of total transmitted information, such as achievable sum rate or system throughput. Yet, BER, which is a crucial aspect of quality of service, is rarely addressed. In our paper we focus on investigating the error performance, which is estimated through Monte Carlo simulation. 

The main contribution of our work is as follows: 

$1)$ We present a RIS aided multi-way communication system model, and analyze its throughput.

 $2)$ We compare three different methodologies for SIC sorting. Including two SIC ordering methods with fixed RIS reflective phases and one SIC ordering method based on RIS phase optimization. 

$3)$ We conduct comprehensive investigations of the BER as a function of the number of users in the system, and the numbers of RIS reflective elements. 

$4)$ We use a previously proposed amplitude-phase relation forf RIS reflection elemnts to modify the  SIC ordering optimization algorithm originally used for constant amplitude elements and evaluate BER performance for this case.

The rest of this paper is organized as follows. Section \ref{model_multiway} presents the background and a model for the multi-way communication system. Performance evaluation of BER for three SIC ordering methods is presented in section \ref{per_constampl}, for various numbers of reflection elements and users when the magnitudes of RIS refelction coefficients are constant. The case of amplitude variations in  RIS reflection coefficinets is considered in section \ref{mag_var}. The couclusions of this paper are provided in section \ref{concl}.

\enlargethispage*{3mm}

\section{RIS aided Multi-Way communication and SIC Methods} \label{model_multiway}

\subsection{ A Model for Multi-Way Communication and Throughput Considerations}

Consider a multi-way communication system where $K$ users exchange information aided by $L$ relays as in \cite{hou2020network}. The operation of such a system, which is based on a time division usage of the channel,  can be summarized in two stages. During stage 1, each user, using its own time slot, broadcasts its information symbol to the other $K-1$ users and $L$ relays in a dedicated time slot. Hence stage 1 consists of $K$ channel uses. During stage 2 each relay broadcasts its network coded information symbol in a dedicated time slot. Hence stage 2 consumes $L$ channel uses.
Such a multi-way communication system requires $K+L$ channel uses to exchange $K$ information symbols, resulting in a throughput of $K/(K+L)$ symbols/channel-uses, which is limited to less than one. This limitation is due to use of time divsion transmission in the two stages. 

To adress this throughput limitation, in this paper we consider a full-duplex based multi-way system where transmission from all users and reception is done simulataneusly, without the use of relays. Instead we employ RIS technology and exploit its full duplex capabilities, allowing to transmit and receive all $K$ information symbols in one channel use \cite{li2020intelligent} \cite{liu2026enabling} \cite{perveen2025}. Such a scheme  could increase the throughput to $K$ symbols/channel-uses enabling in principle  a linear scaling with the number of users. However, unlike  a time division based system, our RIS fulll-duplex technique is plagued by user interference, which also increases when the number of users is increased. Hence to be succesful, such a system must employ effective interfernce counter measures, and our paper focusses  on this issue. In our system, each user employs a single transmit, and multiple  receive antennas aiding to mitigate user interference during the reception process. Furthermore the number of users $K$ is limited by $M$ the number of receive antennas used by each users. To avoid an underdetermined receiving situation creating serious difficulties in user separation we must have
\vspace*{-4mm}
\begin{equation}
M\ge K-1
\vspace*{-5mm}
\end{equation} 
which takes into account that the transmission from the receiving user does not have to be detected, and hence the detection algorithm has to cope with only $K-1$ transmitting ueers. In such a system the throughput is limited to $M+1$ symbols/channel-use, and the maximal number of users  is $M+1$.

One can consider a combination of such a full-duplex approach with a time-division approach, where we allow the use of $Q_d$ time slots. In each time slot we have $K$ users operating in full duplex mode. Hence such a system can accomodate $KQ_d$ users and requires $T_d$ channel uses. The throughput of such a combined system is still $K=M+1$ symbols/channel-uses, however the limit on the maximal number of users is removed.  In our work we focus on the full-duplex segment accomodating $K$ users, and investigate the BER that can be achieved.

\subsection{A Full-Duplex RIS Aided Multi-Way Communication System} \label{subII-B}
 Consider an  RIS-aided symbol oriented multi-way communication system, which consists of one RIS and $K$  users, each of which is equipped with $1$ transmit  and $M$ receive antennas. Each user in this ystem receives  information symbols from other $K-1$ users\cite{hou2020network}. We assume an RIS with $N$ passive reflective elements, which enable to  control the phase-shifts of incident EM waves. The users in this  system are assumed to operate in full-duplex mode when transmitting to and receiving from the RIS. Hence all transmissions and receptions are completed during the same time slot. For simplicity, in the present work we assume CSI to be available. 

We consider a very slow fading scenario and hence employ a quasi-static Rayleigh block fading model for all channels. Then during one block transmission, the fading coefficients remain constant. Let the set  $\mathcal{S} =\left \{ T_{1},T_{2},\dots,T_{K}\right \} $ denote all terminal nodes.The channel coefficient from the ${k}$th user's transmit antenna to the ${n}$th RIS element is denoted as $h_{kn}=\alpha_{kn} e^{-j\varphi _{kn}}$, with $\alpha_{k,n} $ and $\varphi _{kn}$, indicating the magnitude and phase respectively of the fading coefficients, where $k\in\left \{ 1,2,...,K \right \}$, $n \in \left \{ 1,\dots ,N \right \}$. Similarly, the channel coefficient from the ${n}$th RIS element to the ${k}$th user's $m$th receive antenna is denoted as $h_{knm}=\alpha_{knm} e^{-j\varphi _{knm}}$, where $\alpha_{knm} $ and $\varphi _{knm}$ indicate the magnitude and phase respectively, $k\in\left \{ 1,2,...,K \right \}$, $m\in\left \{ 1,2,...,M \right \}$, $n \in \left \{ 1,\dots ,N \right \}$. Since the magnitudes of channel fading coefficients are assumed to be independently Rayleigh distributed, and the corresponding phases are uniformly distributed in $[ 0 , 2\pi )$, then $h_{kn}$ and $h_{knm}$ are complex Gaussian random variables with zero mean and variance $\zeta_k$, $\zeta_{km}$ respectively. Furthermore, $\mathbf{h}_{k}=
\begin{bmatrix}
	{h}_{k1}&  {h}_{k2}&  \dots&{h}_{kN}
\end{bmatrix}^{T}\in \mathbb{C}^{N\times1}$ indicates the channel vector between the $k$th transmitting user and the RIS, where $\mathbb{C}$ is the complex domain, $\mathbf{h}_{km}=
\begin{bmatrix}
	{h}_{k1m}&  {h}_{k2m}&  \dots&{h}_{kNm}
\end{bmatrix}^{T}\in \mathbb{C}^{N\times1}$ indicates the channel vector between the RIS and the ${m}$th receive antenna of user $k$, $\mathbf{H}_{kM}=
\begin{bmatrix}
	\mathbf{h}_{k1}&  \mathbf{h}_{k2}&  \dots&\mathbf{h}_{kM}
\end{bmatrix}\in \mathbb{C}^{N\times M}$ indicates the channel matrix between the RIS and the $k$th receiving user. The system is illustrated in Figure $1$.
\begin{figure}[H]
\centering
\scalebox{0.8} {
	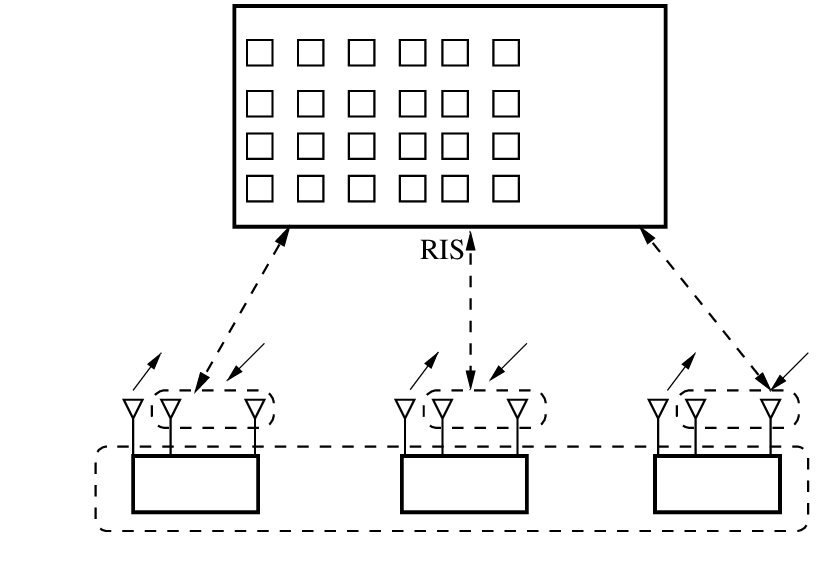 
	}
	\caption{A model for an RIS aided multi-way communication system}
	\label{1}
\end{figure}
Assume equal power allocation with $P_k=P_{all}/K$ for $k \in \left\{1,2,\dots, K \right\}$, where $P_{all}$ is the total power allocated to the network and $P_k$ is the transmit power of the $k$th user. Let  $\mathbf{y}_{kM}=
\begin{bmatrix}
	{y}_{k1}&  {y}_{k2}&  \dots&{y}_{kM}
\end{bmatrix}^{T}\in \mathbb{C}^{M\times1} $indicate the signal received by the $k$th user on all $M$ receive antennas. 
Hence the signal at the $m$th receive antenna of user $k$ can be written as:
\begin{equation}
	\begin{aligned}
		y_{km}=&\sum_{k^{'}=1, k^{'}\ne k }^{K} \sqrt{P_{k^{'}}} (\sum_{n=1}^{N}h_{knm}\eta _{n}e^{j\theta_{n}}h_{k^{'}n})s_{k^{'}}+\underbrace{\sqrt{P_{k}} (\sum_{n=1}^{N}h_{knm}\eta _{n}e^{j\theta_{n}}h_{kn})s_{k}}_{\text{self-interference}}+\underbrace{i_{km}}_{\text{resid. loop-interf.}}+\underbrace{w_{km}}_{\text{AWGN}}\\
		&m\in \left \{ 1,2,\dots ,M \right \} 
	\end{aligned}
	\label{eq:1}
\end{equation}	
where $s_{k^{'}}$ is the transmitted channel symbol of $k^{'}$th user, satisfying $\mathit{E} \left \{ \left | s_{k^{'} }  \right | ^{2} \right \}=E_s =1 $. The first term in (\ref{eq:1}) represents the signals from other $K-1$ users received via RIS reflections. The second term represents self-interference, which is the signal received at the $k$ user from its own transmitted signal via RIS reflections. The third term $i_{km}$ is the residual loop-interference at the $k$th user's $m$th receive antenna due to full-duplex operation after self interference cancellation as in \cite{li2020intelligent}. According to \cite{rodriguez2014performance},  imperfect knowledge of the transmitted signal leads to  imperfections in the interference cancellation process. The residual loop-interference is zero-mean, additive  Gaussian\cite{rodriguez2014performance}. We model $i_{km}$ as complex Gaussian   $i _{km}\sim \mathcal{CN} (0,\sigma_{i_{km} } ^{2})$. .The noise $w _{km}$ is AWGN  $w _{km}\sim \mathcal{CN} (0,\sigma_{w_{km} } ^{2})$.
 In vector form (\ref{eq:1}) can be expressed as:
\begin{equation}
	\begin{aligned}
		&y_{km}=\sum_{k^{'}=1, k^{'}\ne k }^{K} \sqrt{P_{k^{'}}}\begin{bmatrix}
			{h}_{k1m}& {h}_{k2m} & \dots &{h}_{kNm}
		\end{bmatrix} \begin{bmatrix}
			\eta _{1}e^{j\theta_{1}}&  &  & \\
			&  \eta _{2}e^{j\theta_{2}}&  & \\
			&  &  \ddots & \\
			&  &  &\eta _{N}e^{j\theta_{N}}
		\end{bmatrix}                  \begin{bmatrix}
			{h}_{k^{'}1}\\
			{h}_{k^{'}2}\\
			\vdots \\
			{h}_{k^{'}N}
		\end{bmatrix}s_{k^{'}}\\
		&+\sqrt{P_{k}}\begin{bmatrix}
			{h}_{k1m}& {h}_{k2m} & \dots &{h}_{kNm}
		\end{bmatrix} \begin{bmatrix}
			\eta _{1}e^{j\theta_{1}}&  &  & \\
			&  \eta _{2}e^{j\theta_{2}}&  & \\
			&  &  \ddots & \\
			&  &  &\eta _{N}e^{j\theta_{N}}
		\end{bmatrix}                     \begin{bmatrix}
			{h}_{k1}\\
			{h}_{k2}\\
			\vdots \\
			{h}_{kN}
		\end{bmatrix}s_{k}+n_{km}
	\end{aligned}
\end{equation}	  
where for simplicity we use $n_{km}=w_{km} + i_{km}$ to denote the aggregrate noise term.
When the $k$th user aims to decode the signal from the $j$th user, (\ref{eq:1}) can be rewritten as:
\begin{equation}
	\begin{aligned}
		&y_{km}=\underbrace{\sqrt{P_{j} }(\mathbf{h}_{km})^{T} \mathbf{\Theta}\mathbf{h}_{j}s_{j}}_{\text{desired ${j}$th transmitter signal}}+\underbrace{\sqrt{P_{k} }(\mathbf{h}_{km})^{T} \mathbf{\Theta}\mathbf{h}_{k}s_{k}}_{\text{self-interference}}+ \sum_{k'=1,k'\neq k,j}^{K}\underbrace{\sqrt{P_{k'} }(\mathbf{h}_{km})^{T} \mathbf{\Theta}\mathbf{h}_{k'}s_{k'}}_{\text{inter-user interference}}+ n_{km},\\
		&m\in \left \{ 1,2,\dots ,M \right \} 
	\end{aligned}
\end{equation}
where $\mathbf{\Theta}=\mathrm{diag}(\eta _{1}e^{j\theta _{1}},\dots,\eta_{n}e^{j\theta_{n}},\dots,\eta _{N}e^{j\theta_{N}})$ indicates a diagonal matrix representing the reflective properties of the RIS, 
\vspace*{-4mm}
\begin{equation}
\mathbf{n}_{k}=
\begin{bmatrix}
	{n}_{k1}&  {n}_{k2}&  \dots&{n}_{kM}
\end{bmatrix}^{T}\in \mathbb{C}^{M\times 1}. 
\vspace*{-5mm}
\end{equation}
Then the received signal vector of the $k$th user 
\vspace*{-5mm}
\begin{equation}
\mathbf{y}_{kM} =
\begin{bmatrix}
	{y}_{k1}&  {y}_{k2}&  \dots&{y}_{kM}
\end{bmatrix}^{T}\in \mathbb{C}^{M\times1}, 
\vspace*{-6mm}
\end{equation}
can be expressed as:
\vspace*{-5mm}
\begin{equation}
	\mathbf{y}_{kM}=\underbrace{\sqrt{P_{j} }(\mathbf{H}_{kM})^{T} \mathbf{\Theta}\mathbf{h}_{j}s_{j}}_{\text{desired ${j}$th transmitter signal}}+\underbrace{\sqrt{P_{k} }(\mathbf{H}_{kM})^{T} \mathbf{\Theta}\mathbf{h}_{k}s_{k}}_{\text{self-interference}}+\sum_{k'=1,k'\neq k,j}^{K}\underbrace{\sqrt{P_{k'} }(\mathbf{H}_{kM})^{T} \mathbf{\Theta}\mathbf{h}_{k'}s_{k'}}_{\text{inter-user interference}}+\mathbf{n}_{k}
	\label{eq:received}
	\vspace*{-3mm}
\end{equation}
We assume the components of the fading matrix $\mathbf{H}_{kM}$ follow a zero mean  complex Gaussian  distribution  with  uncorrelated components  \cite{semmler2023high}\cite{semmler2023zero}\cite{guo2020differential}., as described in subsection \ref{subII-B}.

For decoding in our multi-way RIS system,  the $j$th user receiver knows its own transmitted symbol, thereby enabling the mitigation of self-interference in (\ref{eq:received}) prior to the decoding of signals from the other users\cite{li2020intelligent}. Furthermore, the term corresponding to inter-user interference in (\ref{eq:received}) represents the contributions from all transmitted signals except those from the $j$th and $k$th user. We use  SIC decoding\cite{li2020intelligent} to cope with such a situation.

\enlargethispage*{5mm}
\subsection{SIC Techniques for RIS aided multi-way communications}
We employ ordered SIC, where decoding progresses sequentially according to some ordering procedure. During each SIC decoding iteration, the $k$th user decodes a single symbol from an interfering user. The detailed SIC algorithm  is presented in \cite{patel1994analysis} \cite{wolniansky1998v} \cite{lai2021reconfigurable}. The strongest user according to a suitable criterion is decoded  first. Then its interference contribution is re-constructed  and subtracted from the received signals. In the second step, the strongest of the remaining users is decoded and cancelled in the same fashion. This procedure repeats until the weakest user is decoded. There are many variations to the basic SIC algorithm, and in this work we consider three which seem to be more suitable for our application.
In our work  we start with the minimum mean squared error  (MMSE) SIC detection technique \cite{choi2010optimal}, rather than zero forcing  (ZF) SIC\cite{choi2010optimal}. 

 Assume the $k$th user receives symbols from other $K-1$ users, and  $r_{k} (n) $ represents the  decoded user in the $n$th SIC step at the $k$th receiver. It is noteworthy to mention that the $k$th user will not decode its own signal, hence, $1\le n \le K-1$ and $1\le r_{k} (n)\le K$ with $r_{k} (n) \neq k$. 
Before the $n$th step, the remaining symbols to be decoded at the $k$th user receiver are those from  users $r_{k} (n), \dots, r_{k} (K-1) $. We use $\mathbf{y}_{kM}^{\left ( n \right ) }  $ to represent the remaining received components  from $\mathbf{y}_{kM}$ in (\ref{eq:received}) at the start of iteration $n$ which can be expressed as
\begin{equation}
	\begin{aligned}
		\mathbf{y}_{kM}^{\left ( n \right ) }=\sum_{m=n}^{K-1}\sqrt{P_{r_{k} (m)} }(\mathbf{H}_{kM})^{T} \mathbf{\Theta}\mathbf{h}_{r_{k} (m)}s_{r_{k} (m)}+\mathbf{n}_{k}
		=\mathbf{Gs}+\mathbf{n}_{k}
	\end{aligned}
\end{equation} 
where $\mathbf{G}=\left [ \sqrt{P_{r_{k} (n)} }(\mathbf{H}_{kM})^{T}\mathbf{\Theta}\mathbf{h}_{r_{k} (n)}, \dots,  \sqrt{P_{r_{k} (K-1)} }(\mathbf{H}_{kM})^{T}\mathbf{\Theta}\mathbf{h}_{r_{k} (K-1)} \right ]=\left [ \mathbf{g}_{r_{k} (n)}, \dots, \mathbf{g}_{r_{k} (K-1)} \right ]$ represents the corresponding  channel vectors, and $\mathbf{s}=
\begin{bmatrix}
	s_{r_{k}(n)} &  s_{r_{k}(n+1)} & \dots &  s_{r_{k}(K-1)}
\end{bmatrix}$ represents the corresponding symbols. According to \cite{choi2010optimal}, the MMSE weight vector is used to suppress interference in detecting a symbol from the received signal vector. Let $\mathbf{w}_{k}^{\left ( n \right ) }$ denote the weight vector that minimizes the mean squared error at the $n$th step of user $k$. 
Defining $\mathbf{w}$ to be an $K-n$ dimensional vector, then  the  weight vector $\mathbf{w}_{k}^{\left ( n \right ) }$ used for detecting  $s_{r_{k}(n)}$ from $\mathbf{y}_{kM}^{\left ( n \right ) }$ is given by
\begin{equation}
	\mathbf{w}_{k}^{\left ( n \right ) }=\arg \min_{\mathbf{w}}\mathit{E} \left [ \left | s_{r_{k}(n)}-\mathbf{w}^H\mathbf{y}_{kM}^{\left ( n \right ) } \right |^2  \right ] 
\end{equation} 
Next we derive a closed form expression for $\mathbf{w}_{k}^{\left ( n \right ) }$. Define the objective function\\ $L=\mathit{E} \left [ \left | s_{r_{k}(n)}-\mathbf{w}^H\mathbf{y}_{kM}^{\left ( n \right ) } \right |^2  \right ]$. Since $L$ is quadratic with respect to $\mathbf{w}$,  the optimal solution is obtained when $\frac{\partial L}{\partial \mathbf{w} } =0$. Here $\mathbf{w} $ is a complex vector, $\frac{\partial L}{\partial \mathbf{w} }$ denotes the complex gradient with respect to $\mathbf{w}$. Based on the derivate of a real-value function with respect to a complex variable\cite{palomar2005gradient} \cite{brandwood1983complex}, we get:

\begin{equation}
		\mathbf{w}_{k}^{\left ( n \right ) }
		=\left (\mathbf{G}\mathbf{G}^{H}+\frac{ N_0 \mathbf{I}}{E_s}\right )^{-1}\mathbf{g}_{r_{k}(n)}
	\label{eq:nth_weight}
\end{equation}

In the initial step, the $k$th user will subtract its own signal from the received signal. The result is denoted as $\mathbf{y}_{kM}^{\left ( 1 \right )}=\mathbf{y}_{kM}-\sqrt{P}_{k}(\mathbf{H}_{kM})^{T} \mathbf{\Theta}\mathbf{h}_{k} s_{k}$. Hence at step $n+1$ after cancelling the interference from $n$ users we have  $\mathbf{y}_{kM}^{\left ( n+1 \right )}=\mathbf{y}_{kM}^{\left ( 1 \right )}-\sum_{q=1}^{n}\sqrt{P_{r_{k} (q)} }(\mathbf{H}_{kM})^{T} \mathbf{\Theta}\mathbf{h}_{r_{k} (q)} s_{r_{k} (q)}+\mathbf{n}_{k}$. 
If SIC is perfect, we can assume $\hat{s}_{r_{k}(n)}=s_{r_{k}(n)}$, which leads the $k$th user signal in the $(n+1)$th step to be:
\begin{equation}
	\mathbf{y}_{kM}^{\left ( n+1 \right )}=\sum_{q=n+1}^{K-1}\sqrt{P_{r_{k} (q)} }(\mathbf{H}_{kM})^{T} \mathbf{\Theta}\mathbf{h}_{r_{k} (q)}\hat{s}_{r_{k}(q)}+\mathbf{n}_{k}
\end{equation}
Ordering the users for SIC detection affects the system performance. In this paper we consider  three  SIC ordering methods. 

\subsubsection{SIC Ordered by Channel Coefficient Variance}
We firstly choose SIC ordering to be based on the channel variances. The variances of fading coefficients $h_{kn}$   in section \ref{subII-B}, denoted as  $\zeta _{k}$,
capture the path loss of individual transmitters. According to \cite{li2020intelligent} and \cite{silva2020noma}, the channel fading coefficients include range-dependent path-loss, and the user's $r_k\left ( n \right )$ is determined by the descending order of channel fading coefficients variances. From subsection \ref{subII-B} , we have $\mathbf{h}_{k}=\sqrt{\zeta _{k} } \widetilde{\mathbf{h}}_{k}$, $\widetilde{\mathbf{h}}_{k} \sim \mathcal{CN} \left ( 0, \mathbf{I}_{N}\right )$ \cite{silva2020noma}. Hence, the transmitted signals can be organized in a descending order  $\zeta _{1} >\zeta _{2}>\zeta _{3}>\dots>\zeta _{K}$ determiing the detection order.  The practical implementation of this technique is shown in Algorithm 1.

\begin{algorithm}[H]
	%\textsl{}\setstretch{1.8}
	\renewcommand{\algorithmicrequire}{\textbf{Input:}}
	\renewcommand{\algorithmicensure}{\textbf{Output:}}
	\caption{SIC Ordered by Variances}
	\label{alg1}
	\begin{algorithmic}[1]
		\ENSURE  Signal decoded by the $k$th user
		\REQUIRE $n$
		\STATE Initialization:$\left\{ {d_0=10,d,\beta=2.8,N,K,M} \right\}$
		\STATE  Generate the signal received by the $k$th user
		\STATE  Subtract the $k$th user's self-interference
		\FOR{$1\le n\le K-1$}
		\STATE $n \leftarrow n + 1$
		\STATE Sort transmitters by the descending order of $\zeta_k$
		\STATE Compute $\mathbf{w}_{k}^{\left ( n \right ) }$ based on Equation$(10)$
		\STATE Update $\hat{s}_{r_{k}(n)}$ based on hard-decision operation $\hat{s}_{r_{k}(n)}=\mathrm{HD}\left((\mathbf{w}_{k}^{\left ( n \right ) }\right)^{H}\mathbf{y}_{kM}^{\left ( n \right )})$
		\STATE Compute $\mathbf{y}_{r_{k} (n)}$ based on subtract $\hat{s}_K(n)$: $\mathbf{y}_{kM}^{\left ( n+1 \right )}=\mathbf{y}_{kM}^{\left ( n \right )}-\sqrt{P_{r_{k} (n)} }(\mathbf{H}_{kM})^{T} \mathbf{\Theta}\mathbf{h}_{r_{k}(n)}\hat{s}_{r_{k}(n)}$	  
		\STATE   Store the Gray coded bits
		\ENDFOR
	\end{algorithmic}  
\end{algorithm}

\subsubsection{SIC Ordered by Least Mean Squared Error}
The second method we consider is SIC ordering  based on the least mean squared error (LMSE) criterion. With this method, the next detected symbol is the one with the smallest mean squared error (MSE)\cite{choi2010optimal}. In the $n$th step, the $k$th user decodes the signal sent by the $r_{k} (n) $th user, and the MSE for each $k$ can be computed using (\ref{eq:nth_weight}) as:
\begin{equation}
	\begin{aligned}
		MSE_k^{\left ( n \right )} &=\mathit{E}\left [ \left | s_{r_{k}(n)}- \left ( \mathbf{w}_{k}^{\left ( n \right ) }\right )^H\mathbf{y}_{kM}^{\left ( n \right )}\right |^2\right ] \\
		&=\mathit{E}\left [ \left ( s_{r_{k}(n)}- \left ( \mathbf{w}_{k}^{\left ( n \right ) }\right )^H\mathbf{y}_{kM}^{\left ( n \right )}\right ) \left ( s_{r_{k}(n)}- \left ( \mathbf{w}_{k}^{\left ( n \right ) }\right )^H\mathbf{y}_{kM}^{\left ( n \right )}\right )^{*} \right ]\\
		&= \mathit{E}\left [ s_{r_{k}(n)}s_{r_{k}(n)}^{*} -  \left ( \mathbf{w}_{k}^{\left ( n \right )}\right )^H\mathbf{y}_{kM}^{\left ( n \right )}s_{r_{k}(n)}^{*} -s_{r_{k}(n)}\left ( \mathbf{y}_{kM}^{\left ( n \right )} \right )^{H} \mathbf{w}_{k}^{\left ( n \right )} + \left ( \mathbf{w}_{k}^{\left ( n \right )}\right )^H\mathbf{y}_{kM}^{\left ( n \right )}\left ( \mathbf{y}_{kM}^{\left ( n \right )}\right )^{H}\mathbf{w}_{k}^{\left ( n \right )} \right ]\\
		&=E_s- \left ( \mathbf{w}_{k}^{\left ( n \right )} \right ) ^H\mathit{E}\left [ \left (\mathbf{Gs}+\mathbf{n}_{k} \right )s_{r_{k}(n)}^{*} \right ]-\mathit{E}\left [ s_{r_{k}(n)} \left ( \mathbf{Gs}+\mathbf{n}_{k} \right )^{H} \right ]\mathbf{w}_{k}^{\left ( n \right )}\\
		&+\left ( \mathbf{w}_{k}^{\left ( n \right )} \right ) ^H \mathit{E}\left [ \mathbf{y}_{kM}^{\left ( n \right )}\left ( \mathbf{y}_{kM}^{(n)}\right )^{H} \right ] \mathbf{w}_{k}^{\left ( n \right )}\\		
		&=E_s- \left ( \mathbf{w}_{k}^{\left ( n \right )} \right ) ^H E_s\mathbf{g}_{r_{k} (n)}-E_s\mathbf{g}_{r_{k} (n)}^{H}\mathbf{w}_{k}^{\left ( n \right )}+\left ( \mathbf{w}_{k}^{\left ( n \right )} \right ) ^H \left ( E_s \mathbf{G}\mathbf{G}^{H}+N_0 \mathbf{I} \right )  \mathbf{w}_{k}^{\left ( n \right )}\\
		&=E_s- \mathbf{g}_{r_{k} (n)}^{H}\left ( \mathbf{G}\mathbf{G}^{H}+\frac{N_0 \mathbf{I}}{E_s}\right )^{-1}  E_s\mathbf{g}_{r_{k} (n)}-E_s\mathbf{g}_{r_{k} (n)}^{H}\left ( \mathbf{G}\mathbf{G}^{H}+\frac{N_0 \mathbf{I}}{E_s}\right )^{-1}\mathbf{g}_{r_{k} (n)}\\
		&+\mathbf{g}_{r_{k} (n)}^{H}\left ( \mathbf{G}\mathbf{G}^{H}+\frac{N_0 \mathbf{I}}{E_s}\right )^{-1} \left ( E_s \mathbf{G}\mathbf{G}^{H}+N_0 \mathbf{I} \right ) \left ( \mathbf{G}\mathbf{G}^{H}+\frac{N_0 \mathbf{I}}{E_s}\right )^{-1}\mathbf{g}_{r_{k} (n)}\\
		&=E_s-E_s \mathbf{g}_{r_{k} (n)}^{H}\left ( \mathbf{G}\mathbf{G}^{H}+\frac{N_0 \mathbf{I}}{E_s}\right )^{-1}\mathbf{g}_{r_{k} (n)}
	\end{aligned}
\end{equation}
Notice that $\mathbf{G}\mathbf{G}^{H}+\frac{N_0 \mathbf{I}}{E_s}$ is Hermitian, so $\left ( \mathbf{G}\mathbf{G}^{H}+\frac{N_0 \mathbf{I}}{E_s}\right )^{H}=\left ( \mathbf{G}\mathbf{G}^{H}+\frac{N_0 \mathbf{I}}{E_s}\right )$.

Let the index of the detected symbol be $r_k(n)$, and the corresponding vector $\mathbf{w}_{k}^{\left ( n \right ) }$. We can thus find the symbols to be detected:
\begin{equation}
	\begin{aligned}
		\left \{ r_{k}(n), \mathbf{w}_{k}^{\left ( n \right ) }\right \}&= \arg \min_{l,\mathbf{w}}\mathit{E}{[|s_{l}-\mathbf{w}^{H}\mathbf{y}_{kM}^{(n)}|^2]}\\
		&=\arg\min_{l}(E_S-E_S\mathbf{g}_{l}^H(\mathbf{G}\mathbf{G}^H+\frac{N_0}{E_S}\mathbf{I})^{-1}\mathbf{g}_{l})
	\end{aligned}
\end{equation}
Here we summarize the procedure of ordered LMSE SIC: The first detected symbol is $\left \{ r_{k}(1), \mathbf{w}_{k}^{\left ( 1 \right ) } \right \}= \mathop{\arg\min}\limits_{l \ \mathbf{w}}E{[|s_{l}-\mathbf{w}^{H}\mathbf{y}_{kM}^{(1)}|^2]}$.  We then subtract the interference of $s_{r_{k}(1)} $ to get $\mathbf{y}_{kM}^{(2)}=\mathbf{y}_{kM}^{(1)}-\mathbf{g}_{r_{k}(1)} \widehat{s}_{r_{k}(1)}$, where $\widehat{s}_{r_{k}(1)}$ is a hard decision of $s_{r_{k}(1)}$ based on $\mathbf{w}_{r_k(1)}^H\mathbf{y}_{kM}^{(1)}$. Furthermore, we select the user with the smallest MSE among the remaining user symbols to be the second detected user: $\left \{ r_{k}(2), \mathbf{w}_{r_{k}(2)} \right \}= \mathop{\arg\min}\limits_{l\in I_1 \ \mathbf{w}}E{[|s_{l}-\mathbf{w}^{H}\mathbf{y}_{kM}^{(2)}|^2]}$, where $I_1=I_0\setminus r_k(1), I_0=\left \{ 1,2,...,k-1,k+1, \dots, K \right \} $. This LMSE combining and cancelling process will be repeated until all symbols are detected. The practical implementation of this algorithm is shown in Algorithm 2.

\begin{algorithm}[H]
	%\textsl{}\setstretch{1.8}
	\renewcommand{\algorithmicrequire}{\textbf{Input:}}
	\renewcommand{\algorithmicensure}{\textbf{Output:}}
	\caption{SIC Ordered by LMSE}
	\label{alg1}
	\begin{algorithmic}[1]
		\ENSURE  Signal decoded by the $k$th user
		\REQUIRE $n$
		\STATE Initialization:$\left\{ {d_0=10,d, \beta=2.8, N, K, M} \right\}$
		\STATE  Generate the signal received by the $k$th user
		\STATE  Subtract the $k$th user's self-interference
		\FOR{$1\le n\le K-1$}
		\STATE $n \leftarrow n + 1$
		\STATE Sort transmitters by LMSE
		\STATE Compute $\mathbf{w}_{k}^{\left ( n \right ) }$ based on Equation$(10)$
		\STATE Update $\hat{s}_{r_{k}(n)}$ based on hard-decision operation $\hat{s}_{r_{k}(n)}=\mathrm{HD}( \left(\mathbf{w}_{k}^{\left ( n \right ) }\right)^{H}\mathbf{y}_{kM}^{\left ( n \right )})$
		\STATE Compute $\mathbf{y}_{r_{k} (n)}$ based on subtract $\hat{s}_K(n)$: $\mathbf{y}_{kM}^{\left ( n+1 \right )}=\mathbf{y}_{kM}^{\left ( n \right )}-\sqrt{P_{r_{k} (n)} }(\mathbf{H}_{kM})^{T} \mathbf{\Theta}\mathbf{h}_{r_{k}(n)}\hat{s}_{r_{k}(n)}$	  
		\STATE   Store the Gray coded bits
		\ENDFOR
	\end{algorithmic}  
\end{algorithm}

\subsubsection{SIC Ordered by RIS Beamforming Optimization}

 Now we consider SIC ordering  based on  beamforming optimization  of RIS phase shift.
Let $\mathbf{H}_K=[\mathbf{h}_1,\mathbf{h}_2, \dots, \mathbf{h}_K] \in \mathbb{C}^{N\times K}$ where as in subsection \ref{subII-B}, $\mathbf{h}_k$ denotes the 
channel vector between user $k$ and the RIS. Furthermore,
$\mathbf{P}=\mathrm{diag} \left ( \sqrt{P_{1}}, \sqrt{P_{2}},\dots, \sqrt{P_{k}} \right ) \in \mathbb{C}^{K\times K}$, $\mathbf{s}=\left [ s_1, s_2, \cdots ,s_k\right ]^{T}   \in \mathbb{C}^{K\times 1}$. Then the received signal at the $k$th user in $(4)$ can be expressed as: $\mathbf{y}_{kM}=\mathbf{H}_{kM}^T\mathbf{\Theta }\mathbf{H}_K\mathbf{P}\mathbf{s}+\mathbf{n}_k $. 
According to \cite{zuo2020resource}, maximizing the combined channel gain by adjusting the RIS reflecting coefficients represented by $\mathbf{\Theta }$ can improve the system throughput. A low complexity decoding order optimization method was introduced in \cite{zuo2020resource}, focusing on maximizing the sum of all the combined channel gains. According to \cite{mu2020exploiting} \cite{qiu2023intelligent}, the user with the highest channel gains will be decoded first. Refering to the framework established in \cite{zuo2020resource}, the optimization problem for our system can be formulated as follows, where the objective function is the  channel gains sum of all users.
\begin{equation}
	\begin{aligned} \label{eq:opt_gains}
		\mathrm{(P1) :} & \max_{\mathbf{\Theta }}\sum_{k=1}^{K}\left \| \mathbf{H}_{kM}^T\mathbf{\Theta }\mathbf{H}_K\mathbf{P} \right \|_F ^{2} \\
		&\begin{array}{l@{\quad}l@{}l@{\quad}l}
			s.t. & 0\le \theta _{n}\le2\pi, \forall  n = 1,2,\dots N. \\
		\end{array}
	\end{aligned}
\end{equation}
where $\left\| \cdot \right\|_F$ denotes the matrix Frobenius norm.
Let $\bm{ \upsilon } =\left [ \upsilon_{1}, \upsilon_{2}, \dots, \upsilon_{N}  \right ] ^{H}$,  where $\upsilon_{n},  \forall  n$, are complex scalars. The constraints in (\ref{eq:opt_gains}) are then equivalent to $\left | \upsilon_n  \right |=1,  \forall  n = 1,2,\dots N$.  Denote $\mathbf{B}_k=\mathbf{H}_K\mathbf{P}, \mathbf{B}_k \in \mathbb{C}^{N\times K} $. Using  $\mathbf{H}_{kM}$ defined in (\ref{eq:received}),we have

\begin{align*}
	\mathbf{H}_{kM}^T\mathbf{\Theta }\mathbf{B}_k&=\begin{bmatrix}
		\mathbf{h}_{k1}^T\\ 
		\mathbf{h}_{k2}^T\\
		\vdots \\
		\mathbf{h}_{kM}^T
	\end{bmatrix}
	\begin{bmatrix}
		e^{j\theta _{1} }  &  &  & \\
		&  e^{j\theta _{2} }  &  & \\
		&  &  \ddots & \\
		&  &  &e^{j\theta _{N} }  
	\end{bmatrix}\mathbf{B}_k
	=\begin{bmatrix}
		\mathbf{h}_{k1}^T\\ 
		\mathbf{h}_{k2}^T\\
		\vdots \\
		\mathbf{h}_{kM}^T
	\end{bmatrix}
	\begin{bmatrix}
		\upsilon_1&  &  & \\
		&  \upsilon_2&  & \\
		&  &  \ddots & \\
		&  &  &\upsilon_N
	\end{bmatrix}\mathbf{B}_k\\
	&=\begin{bmatrix}
		h_{k1,1}\upsilon_1&  h_{k1,2}\upsilon_2&  \dots & h_{k1,N}\upsilon_N \\
		h_{k2,1}\upsilon_1&  h_{k2,2}\upsilon_2&  \dots & h_{k2,N}\upsilon_N\\
		\vdots&  \vdots&  \ddots &\vdots \\
		h_{kM,1}\upsilon_1&  h_{kM,2}\upsilon_2&  \dots & h_{kM,N}\upsilon_N
	\end{bmatrix}\mathbf{B}_k
	=\begin{bmatrix}
		\bm{\upsilon}^H\mathrm{diag}\left ( \mathbf{h}_{k1}^T \right )\mathbf{B}_k \\ 
		\bm{\upsilon}^H\mathrm{diag}\left ( \mathbf{h}_{k2}^T \right )\mathbf{B}_k \\
		\vdots \\
		\bm{\upsilon}^H\mathrm{diag}\left ( \mathbf{h}_{kM}^T \right )\mathbf{B}_k 
	\end{bmatrix}
	=\begin{bmatrix}
		\bm{\upsilon}^H\mathbf{\Phi }_{k1}\\ 
		\bm{\upsilon}^H\mathbf{\Phi }_{k2}\\
		\vdots \\
		\bm{\upsilon}^H\mathbf{\Phi }_{kM}
	\end{bmatrix}
\end{align*}

where $\mathbf{\Phi }_{ki}=\mathrm{diag}\left ( \mathbf{h}_{ki}^T \right )\mathbf{B}_k \in \mathbb{C}^{N\times K} $. Note that since  $\left \| \mathbf{A}_k \right \|_F^2=\mathrm{tr}\left ( \mathbf{A}_k  \mathbf{A}_k^H \right )$, we have
\begin{equation}
	\begin{aligned} \label{eq:tr_norm}
		\left \| \mathbf{H}_{kM}^T\mathbf{\Theta }\mathbf{B}_K  \right \|_F^2&=\mathrm{tr}\left ( \begin{bmatrix}
			\bm{\upsilon}^H\mathbf{\Phi }_{k1}\\ 
			\bm{\upsilon}^H\mathbf{\Phi }_{k2}\\
			\dots \\
			\bm{\upsilon}^H\mathbf{\Phi }_{kM}
		\end{bmatrix}  \begin{bmatrix}
			\mathbf{\Phi }_{k1}^H\bm{\upsilon} & \mathbf{\Phi }_{k2}^H\bm{\upsilon} & \dots  &\mathbf{\Phi }_{kM}^H\bm{\upsilon}
		\end{bmatrix} \right )\\
		&=\bm{\upsilon}^H\mathbf{\Phi }_{k1}\mathbf{\Phi }_{k1}^H\bm{\upsilon}+\bm{\upsilon}^H\mathbf{\Phi }_{k2}\mathbf{\Phi }_{k2}^H\bm{\upsilon}+\cdots+\bm{\upsilon}^H\mathbf{\Phi }_{kM}\mathbf{\Phi }_{kM}^H\bm{\upsilon}
		=\bm{\upsilon}^H\left ( \sum_{m=1}^{M} \mathbf{\Phi }_{km}\mathbf{\Phi }_{km}^H \right )\bm{\upsilon}
	\end{aligned}
\end{equation}
Based on (\ref{eq:tr_norm}), problem $\mathrm{(P1)}$ can be reformulated as $\mathrm{(P2)}$, which is a complex quadratic optimization problem:
\vspace*{-2mm}
\begin{equation}
	\begin{aligned} \label{P}
		\mathrm{(P2) :} & \max_{\bm{\upsilon}}\bm{\upsilon}^H \left (\sum_{k=1}^{K} \sum_{m=1}^{M} \mathbf{\Phi }_{km}\mathbf{\Phi }_{km}^H \right )\bm{\upsilon}\\
		&\begin{array}{l@{\quad}l@{}l@{\quad}l}
			s.t. & \left | \upsilon_n  \right |=1,  \forall  n = 1,2,\dots N \\
		\end{array}
	\end{aligned}
\end{equation}
\enlargethispage*{11mm}
Define $\mathbf{R}= \sum_{k=1}^{K} \sum_{m=1}^{M} \mathbf{\Phi }_{km}\mathbf{\Phi }_{km}^H$. Note that $\bm{\upsilon}^H\mathbf{R}\bm{\upsilon}=\mathrm{tr}\left (\mathbf{R}\bm{\upsilon}\bm{\upsilon}^H \right )$. Let $\bm{\mathit{V}} =\bm{\upsilon}\bm{\upsilon}^H $. We have  $\bm{\mathit{V}}\succeq 0 $ (nonenegative definite)  and $\mathrm{rank}\left (\bm{\mathit{V}}\right )=1 $ \cite{zuo2020resource}. The problem $\mathrm{(P2)}$ can be transformed to the equivalent problem 
\vspace*{-3mm}
\begin{equation}
	\begin{aligned} \label{P}
		\mathrm{(P2.1) :} & \max_{\bm{\mathit{V}} } \mathrm{tr}\left ( \mathbf{R}\bm{\mathit{V}}  \right )\\
		&\begin{array}{l@{\quad}l@{}l@{\quad}l}
			s.t. &  \mathrm{diag}\left (  \bm{\mathit{V}} \right )  = \mathbf{I} \\
			&\mathrm{rank}\left (  \bm{\mathit{V}} \right )  = 1 \\
			& \bm{\mathit{V}}\succeq 0 
		\end{array}
	\end{aligned}
\end{equation}
Since the rank constraint $\mathrm{rank}\left (\bm{\mathit{V}}\right )=1$ is non-convex\cite{luo2010semidefinite}, while the objective function and all other constraints are convex, the problem $\mathrm{(P2.1)}$ is not a convex optimization problemm and therefore it is hard to solve efficiently. However, we can use semi-definite relaxation (SDR)\cite{luo2010semidefinite} to obtain the relaxed optimization problem.  Consequently, problem $\mathrm{(P2.1)}$ is reduced to a semi definite programming (SDP)\cite{luo2010semidefinite} problem:
\begin{equation}
	\begin{aligned} \label{P}
		\mathrm{(P3) :} & \max_{\bm{\mathit{V}} } \mathrm{tr}\left ( \mathbf{R}\bm{\mathit{V}}  \right )  \\
		&\begin{array}{l@{\quad}l@{}l@{\quad}l}
			s.t. &  \bm{\mathit{V}}_{n,n} =1,  \forall  n = 1,2,\dots N \\
			&\bm{\mathit{V}}\succeq 0
		\end{array}
	\end{aligned}
\end{equation}
Problem $\mathrm{(P3)}$ is a convex optimization problem, so we can use numerical optimization solvers (e.g. CVX) to efficiently obtain the optimal solution. Generally, the solution of $\mathrm{(P3)}$ may not satisfy the rank constraint, thus the optimal solution of $\mathrm{(P3)}$ is usually just an approximation to the optimal solution of $\mathrm{(P2)}$\cite{luo2010semidefinite}. 

The algorithm based on SDP is summarized as follows:
\begin{itemize}
	\item [(1)] 
	We first solve $\mathrm{(P3)}$, and use the result of $\mathrm{(P3)}$ as an approximation of $\mathrm{(P2)}$.  Let the optimal solution of $\mathrm{(P3)}$ be $\bm{\mathit{V}}^*$. We firstly use singular value decomposition on $\bm{\mathit{V}}^{*}$ to obtain $\bm{\mathit{V}}^{*}=\mathbf{U\Sigma U}^H $, where $\mathbf{U}$ is a unitary matrix, and $\mathbf{\Sigma}$ is a diagonal matrix. . 
	\item [(2)] 
	Let $r=\mathrm{rank}(\bm{\mathit{V}}^{*})$, and let $\bm{\mathit{V}}^{*}=\sum_{r=1}^{r}\lambda_{i}\mathbf{u}_{i}\mathbf{u}_{i}^{H}$ denote the eigenvalue decomposition of $\mathbf{V}^*$, where $\lambda_{1}\ge\lambda_{2}\ge\dots\ge\lambda_{r}>0$ are the eigenvalues and $\mathbf{u}_{1}, \mathbf{u}_{2},\dots ,\mathbf{u}_{r}\in \mathbb{C}^n$ are the corresponding columns in matrix $\mathbf{U}$. We then use rank-one approximation to $\bm{\mathit{V}}^{*}$ as $\bm{\mathit{V}}_{1}^{*}=\lambda_{1}\mathbf{u}_{1}\mathbf{u}_{1}^{H}$. Define $\widetilde{\bm{\mathit{v}}} =\sqrt{\lambda _{1}}\mathbf{u}_{1} $, then $\widetilde{\bm{\mathit{v}}} \widetilde{\bm{\mathit{v}}}^{H}=\bm{\mathit{V}}_{1}^{*}$. Note that $\mathrm{rank}(\bm{\mathit{V}}_{1}^{*})=1$.
\end{itemize}
If $\mathrm{rank}(\bm{\mathit{V}}^{*})=1$, then $\widetilde{\bm{\mathit{v}}}$ will be a feasible and optimal solution to $\mathrm{(P2)}$. If $\mathrm{rank}(\bm{\mathit{V}}^{*})> 1$, then $\widetilde{\bm{\mathit{v}}}$ is an approximation to the optimal solution to $\mathrm{(P2)}$, such that $\widetilde{\bm{\mathit{v}}} \widetilde{\bm{\mathit{v}}}^{H}$ is the best rank one approximation to $\bm{\mathit{V}}^{*}$. We consider two cases: if $\widetilde{\bm{\mathit{v}}}$ is feasible for $\mathrm{(P2)}$, then we just use $\widehat{\bm{\mathit{v}}} =\widetilde{\bm{\mathit{v}}}$; if $\widetilde{\bm{\mathit{v}}}$ is not feasible to $\mathrm{(P2)}$, we map $\widehat{\bm{\mathit{v}}}$ to a nearby feasible solution to $\mathrm{(P2)}$. In practice, we choose the components of  $\widehat{\bm{\mathit{v}}}$ to be the normalized components of $\widetilde{\bm{\mathit{v}}}$,  $\widehat{\bm{\mathit{v}}}=\left [ \frac{\widetilde{\bm{\mathit{v}}_1}}{\left | \widetilde{\bm{\mathit{v}}_1} \right | }, \frac{\widetilde{\bm{\mathit{v}}_2}}{\left | \widetilde{\bm{\mathit{v}}_2} \right | }, \dots, \frac{\widetilde{\bm{\mathit{v}}_N}}{\left | \widetilde{\bm{\mathit{v}}_N} \right | }  \right ]$. The practical implementation of this algorithm is shown in Algorithm 3.

 \begin{algorithm}[H]
	%\textsl{}\setstretch{1.8}
	\renewcommand{\algorithmicrequire}{\textbf{Input:}}
	\renewcommand{\algorithmicensure}{\textbf{Output:}}
	\caption{SIC Decoding Order Optimization Algorithm}
	\label{alg1}
	\begin{algorithmic}[1]
		\ENSURE  Decoding order $r_k(n)$
		\REQUIRE $n$
		\STATE Initialization:$\left\{ {d_0=10,d, \beta=2.8, N, K, M} \right\}$
		\STATE  Solve problem (P3) to obtain $\bm{\mathit{V}}^*$
		\IF{$\mathrm{rank}(\bm{\mathit{V}}^*)=1$}
		\STATE  Calculate the nonzero singular value $\lambda_{1}$ of matrix $\bm{\mathit{V}}^*$ and its corresponding left-singular vector $\mathbf{u}_1$ via singular value deconposition
		\STATE  Calculate $\mathbf{\Theta}=\mathrm{diag}\left \{\sqrt{\lambda_{1}}*\mathbf{u}_1  \right \} $
		\ELSE
		\STATE Calculate $\mathbf{\Theta}_{l}$ according to rank one SDR\cite{luo2010semidefinite}
		\ENDIF
		\STATE Calculate all combined channel gains $\left \{ \left \| \mathbf{H}_{kM}^T\mathbf{\Theta }\mathbf{H}_K\mathbf{P} \right \|_F ^{2},k\in \left \{ 1,2,\dots K \right \}  \right \}$ and rank them in descending order for each channel
	\end{algorithmic}  
\end{algorithm}

\subsection{Complexity Analysis}
This section considers the complexity of the three ordering algorithms. According to \cite{cormen2022introduction}, running time is a widely used as a measure for analyzing the complexity of algorithms, since it indicates the number of steps executed.  In our paper we use the $\mathcal{O}$ notation to denote the asymptotic running time in terms of the input size.

In Algorithm $1$ and Algorithm $2$, the time complexity mainly comes from matrix multiplications and matrix inverstion\cite{cormen2022introduction}. In phase 1, the initialization  only requires a constant time, so the time complexity is $\mathcal{O}(1)$. In phase 2, generating input signals needs $\mathcal{O}(K)$ steps; generating channel realizations needs $\mathcal{O}(NK)$ steps; generating RIS coefficients needs $\mathcal{O}(N^2)$ steps; and computing $\mathbf{F}=\mathbf{H}_{kM}^T\mathbf{\Theta }\mathbf{H}_K$ needs $\mathcal{O}(MN^2+MNK)$ steps. The signal received by the $k$th user needs $\mathcal{O}(MK+M)$ steps. Thus the overall time complexity for step 2 is $\mathcal{O}(MN^2+MNK)$ if we keep the highest order terms. In phase 3, subtracting self-interference needs $\mathcal{O}(K)$ steps. From phase 4 to phase 10 for each $n$, computing weight vectors needs $\mathcal{O}\left ( \left ( K-n \right )M\left ( K-n \right )+\left ( K-n \right )^{3}   \right ) $ steps; selecting the index of the next detected symbol requires $\mathcal{O}\left ( K-n \right)$ steps; hard-decoding the selected symbol needs $\mathcal{O}\left ( \left ( K-n \right)M\right)$ steps; cancellation of detected symbol interference neads  $\mathcal{O}\left ( M \right)$ steps. Hence, for $1\le n\le K-1$, and keeping the highest order terms, we have the total  complexity: $\mathcal{O}\left ( \sum_{n=1}^{K-1}\left ( \left ( K-n \right ) M \left ( K-n \right ) +\left ( K-n \right ) ^{3} \right )   \right)=\mathcal{O}\left ( K^{3} M+  K^{4} \right ) $. Therefore, the total time complexity is $\mathcal{O}\left ( MN^{2} +MNK+MK^{3}+ K^{4}\right )$.

In Algorithm 3, the most time consuming operations are computing $\mathbf{R}$ in $\mathrm{(P3)}$ and using SDP to solve $\bm{\mathit{V}}$. In phase 1, the time complexity is $\mathcal{O}(1)$. In phase 2, there is a $K$-loop to compute $\mathbf{R}$. For each user $k$, the complexity of computing $\mathbf{R}$ is $\mathcal{O}\left ( \sum_{i=1}^{M}(NK^2 +N^2K+NK) \right )$. Thus, the total time complexity for this phase is $\mathcal{O}\left ( MNK^2+N^2MK \right )$. Refering to the complexity in \cite{zuo2020resource}, the complexity for solving $\bm{\mathit{V}}$ from phase $3$ to phase $8$ is $\mathcal{O}\left ( (N+1)^6  \right )$ for SDP, and $\mathcal{O}\left ( N^3  \right )$ for singular value decomposition. In phase $9$, the time complexity is altoghther $\mathcal{O}\left ( MN^2+MNK+K\log{K}  \right ) $\cite{harvey2021integer}. Suming  up keeping the highest order terms gives  the  complexity as $\mathcal{O}((N+1)^6+MNK^2+MN^2K)$. These results indicate that for Algorithms 1 and 2, the complexity as a function of the number of users $K$ behaves as $K^4$ and as a function of the number of RIS elements $N$ it behaves as $N^2$. For algorithm 3, the complexity in terms of the number of users behaves as $K^2$ and in terms of the number of RIS elemnts it behaves as $N^4$.

\section{Performance of SIC Methods for RIS with Constant Amplitude Elements (Ideal Case)}\label{per_constampl}

\subsection{Computer simulations setup
}
We consider a Rayleigh quasi-static fading channel where the  fading coefficients are independently distributed, making  $h_{knm}$, and $h_{kn}$ complex Gaussian random variables with zero mean and variance $\zeta_{km}$, $\zeta_k$ respectively. These variances are determined by the path loss for each user given by $10\times (\frac{d}{d_0}) ^{-\beta }$ \cite{li2020relay}.
In our work we employ a reference distance $d_0 =10$, and path loss exponent $\beta=2.8$. Furthermore the users''s distances are $d=100, 110, 120, 130, ...$.
We use QPSK with Gray coding., and equal transmit power for each user, $P_k =2$.  Using Monte-Carlo simulations the average BER over all the received data from $K-1$ users    is calculated for each $E_b/N_0$, based on at least 500-bit errors and 1000 channel realiztions.  The final BER for each $E_B/N_0$ value  is the average of the results over  all channel realizations.We assume that self-intereerence is perfectly cancelled for each user befor decoding. Furthermore, for a system with of up to $K$ users, we employ $M=K-1$ receive antennas, since self interference of the user whose performance is assessed  is cancelled. With one transmitting  user the system degenerates to a single point to point case, where user 1 receives (but not transmits) and user 2 transmits. We consider this case for calibration purposes. With two users, we assume that users 1 and 2  transmit, hence at user 1 receiver there is self-interference cancellation. With perfect self-interference  cancellation this is equivalent to the previous case. For $K>2$ we have that after self interfernce cancellation at user 1 his receiver has to extract information from $K-1 >1$ transmitting users, and hence SIC has to be performed.

\subsection{Results for Four Receive Antennas Per User}

\begin{figure}[thbp!] 
	\centering
	%\begin{minipage}[t]{0.49\linewidth}
		\centering
 	    \includegraphics[scale=0.6]{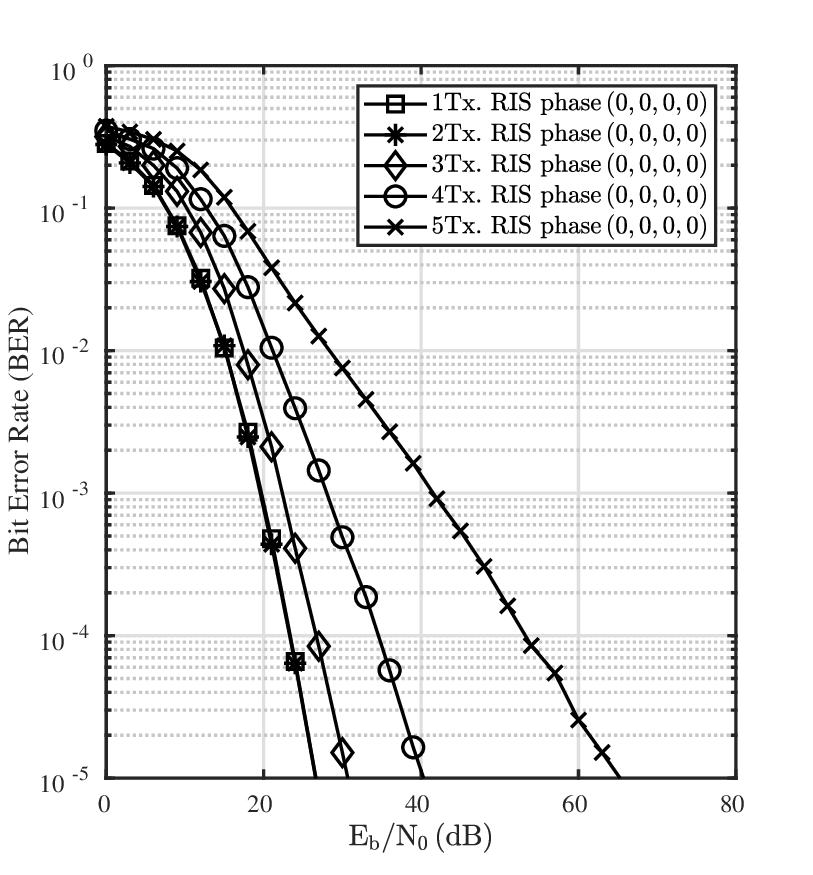}
	    \caption{BER results for $1$ to $5$ transmitting users each employing 4 receive antennas with QPSK, and SIC ordering by channel coefficient variances.}
	    \label{fig2}
   % \end{minipage}
   \end{figure}

\begin{figure}
  %  \begin{minipage}[t]{0.49\linewidth}
	   \centering
	   \includegraphics[scale=0.6]{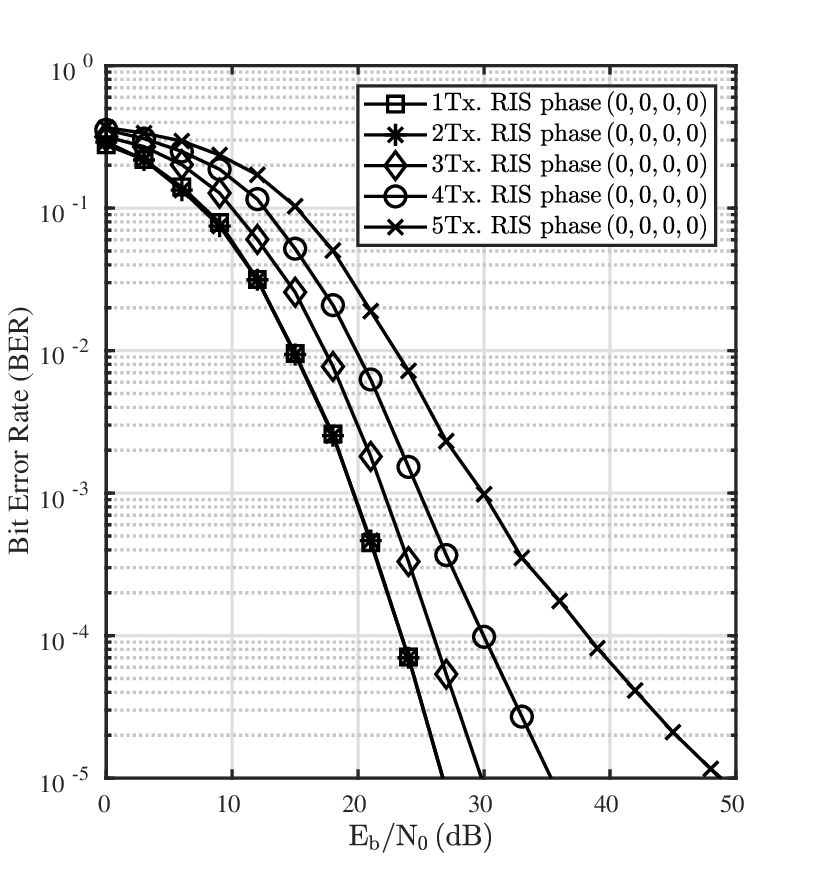}
    	\caption{BER results for $1$ to $5$ transmitting users each employing 4 receive antennas with QPSK and SIC ordering by LMSE.}
    	\label{fig3}
	 % \end{minipage}
\end{figure}

We consider the  scenario where each user is equipped with $M=4$ receive antennas, and the RIS is ideal, meaning that its elements affect the electromagnetic wave through phase shifts only, without any magnitude variations. We start with two SIC ordering methods  represented by algorithms 1 and 2. In Figures \ref{fig2} and \ref{fig3} we present simulation results with four receive antennas per user, and systems consisting of  1 - 5 users with QPSK modulation. The parameters of RIS elements are set to  $N=4$, and $\left ( \theta_{1} ,\theta_{2}, \theta_{3}, \theta_{4}\right )=\left ( 0,0 ,0,0  \right ) $.

In Figure \ref{fig2} we present results for SIC ordering by channel coefficients variance. We see that  with  5 transmitting users  we have a performance loss of $29.8$ dB at a BER of $10^{-4}$ relative to 1 transmitter. With  4 transmitting users, the performance  loss is $12.6$ db relative to a single user system. For a 3 transmitting users system the degradation is reduced to $3$ db.  Next we consider the diversity order.
We see that at a BER of  $10^{-4}$, the achievable diversity order with two transmitting users is 3, for 3 transmitting users it is 2.63 , and with 5 transmitting users it is 0.83.
As a benchmark we also present the asymtotic diversity order following \cite{prasad2004analysis}, which for our system is given by $M-(K-1)+1=M-K+2$.
Numerical results are summarized in Table \ref{table_div_order_4antennas}.

In Figure \ref{fig3} we present BER results for LMSE SIC ordering. We see that  with  5 transmitting users we have  a performance loss of $14.8$ dB at a BER of $10^{-4}$ with respect to the 1 transmitting user case .  The performance loss with  4 transmitting users  is $6.6$ dB,  and with  3 transmitting users it is $2.6$ dB.
Now we consider the diversity provided by LMSE SIC ordering.
With two transmitting users, the achivable diversity order is similar to SIC ordering by channel coefficients variance.
With 3 transmitting users the system provides an achievable diversity gain of $2.63$ at a BER of  $10^{-4}$ , and with 5 transmitting users the achievable diversity order is $0.94$. Notice the clear decrease in the BER curve slope for 5 transmitting users starting at the BER of $3\cdot 10^{-4}$ as a result of the system with 4 receive antennas having to cope with 4 users transmitting simultaneously, a phenomemon well known in MIMO communication systems.
Further numerical results are summarized in Table \ref{table_div_order_4antennas}.

A diversity performance  comparison between Algorithms 1 and  2 is presented in Table \ref{table_div_order_4antennas}. The achievable diversity when SIC is ordered by LMSE is $0.06$ larger than when ordered by the variance with $4$ transmitters. With 5 transmitters, the achievable diversity order for SIC ordered by LMSE is $0.11$ larger than when ordered by variance. From Table \ref{table_div_order_4antennas},  we can see that the achievable diversity is smaller than the asymptotic diversity, but reasonable close  for the BER in the range of $10^{-5}$ as considered in this paper

\begin{table}[H]
	\centering
	\resizebox{\linewidth}{!}{
		\begin{tabular}{|c|c|c|c|}
			\hline
			Number of transmitters &\thead{Achievable diversity\\ (Ordering by variance)}& \thead{Achievable diversity\\(Ordering by LMSE)}& Asymptotic diversity\\ \hline
			3 & 2.63&2.63 & 3\\ \hline
			4 & 1.79&1.85 & 2 \\ \hline
			5 & 0.83&0.94 & 1 \\ \hline
	\end{tabular}}
	\caption{Diversity performance with for four receive antennas per user}
	\label{table_div_order_4antennas}
\end{table}

\subsection{Comparison of the Three SIC Ordering Approaches}
To obtain the numerical results of SIC beamforming optimization (DOO), we initially generate a number of fading channels, then RIS phase shift optimization is applied for each channel realization, followed by  Monte Carlo simulations  conducted for: SIC beamforming with  Algorithm 3.  The BER results for RIS with and without beamforming  optimization are considered in the next subsections.

\subsubsection{Performance for Three Users System}
\begin{figure}[thbp!] 
	\centering
		\includegraphics[scale=0.6]{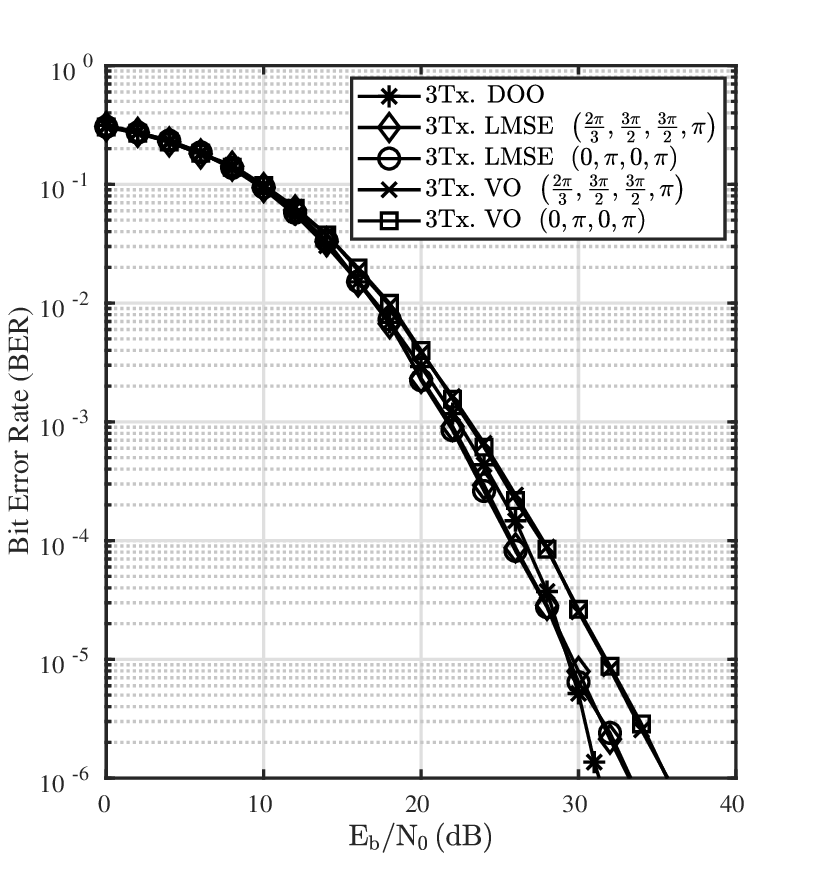}
		\caption{BER of $3$ user systems with QPSK  for three different SIC ordering methods. System parameters: $K=3, N=4, M=4,d=[100,110,120]$m. }
\label{three_users}
	\end{figure}

We first simulate the SIC decoding order optimization algorithm for a three-user system with QPSK modulation, and compare the results with SIC based on variance (VO)  and LMSE orderings. We assume the number of RIS reflecting elements is $N=4$. Two types of RIS phases are assumed for each SIC ordering method : $\left ( \theta_{1} ,\theta_{2}, \theta_{3}, \theta_{4}\right )=\left ( 0,\pi ,0,\pi  \right ) $ and $\left ( \theta_{1} ,\theta_{2}, \theta_{3}, \theta_{4}\right )=\left ( \frac{2\pi }{3} ,\frac{3\pi }{2} ,\frac{3\pi }{2},\pi  \right ) $. The simulation results  depicted in Figure \ref{three_users} show a performance gain of "3Tx. LMSE $\left ( \frac{2\pi }{3} ,\frac{3\pi }{2} ,\frac{3\pi }{2},\pi  \right )$" compared with  "3Tx. LMSE $\left ( 0,\pi ,0,\pi  \right ) $" of around $1$dB at a BER of $10^{-6}$. Similarly, the performance gain of "3Tx. VO $\left ( \frac{2\pi }{3} ,\frac{3\pi }{2} ,\frac{3\pi }{2},\pi  \right )$" compared with  "3Tx. VO $\left ( 0,\pi ,0,\pi  \right ) $" is around $2$dB at a BER of $10^{-6}$.
Beamformimg optimization starts to perform better than the other two ordering methods only at BER$<7\cdot10^{-6}$.

	\begin{figure}[thbp]
		\centering
		\includegraphics[scale=0.59]{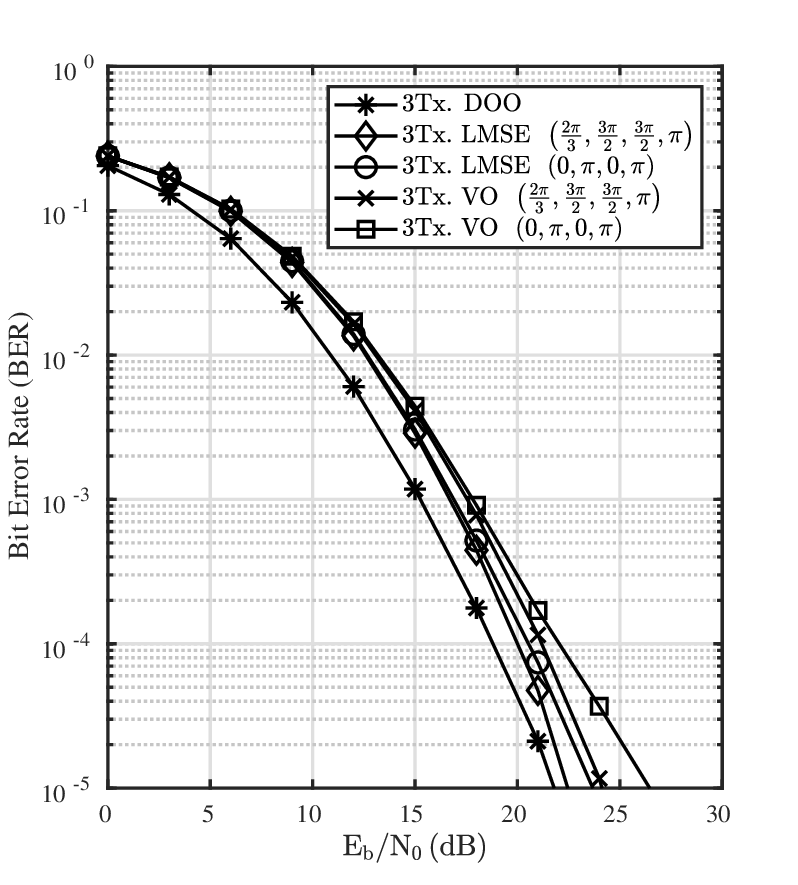}
		\caption{BER  of $3$ user systems with QPSK for three SIC Ordering Methods. System parameters: $K=3, N=8, M=4,d=[100,110,120]$m. The RIS phases are grouped in sets of four, with each set being $\left ( \frac{2\pi }{3} ,\frac{3\pi }{2}, \frac{3\pi }{2}, \pi \right ) $ or $\left ( 0 ,\pi, 0, \pi \right ) $}
\label{N=8}
\end{figure}

\begin{figure}[thbp!] 
	\centering
		\centering
		\includegraphics[scale=0.6]{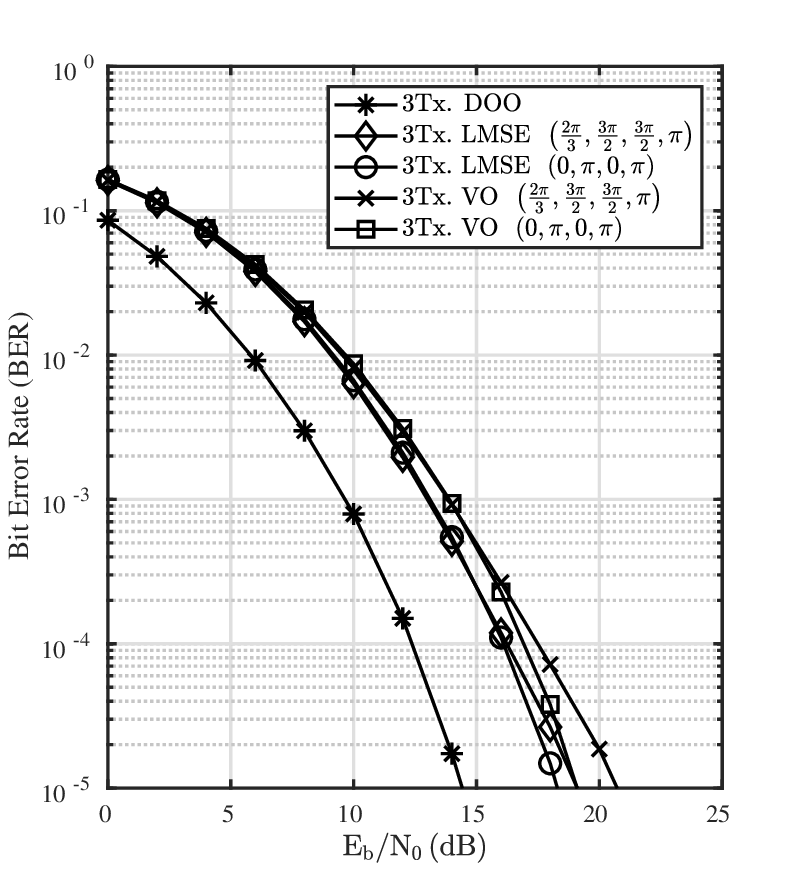}
		\caption{BER results for a $3$ user system with QPSK, and three  SIC ordering methods. System parameters: $K=3, N=16, M=4,d=[100,110,120]$m. The RIS phases are grouped in sets of four, with each set being $\left ( \frac{2\pi }{3} ,\frac{3\pi }{2}, \frac{3\pi }{2}, \pi \right ) $ or $\left ( 0 ,\pi, 0, \pi \right ) $}
		\label{N=16}
	\end{figure}

\begin{figure}[thbp!]
		\centering
		\includegraphics[scale=0.6]{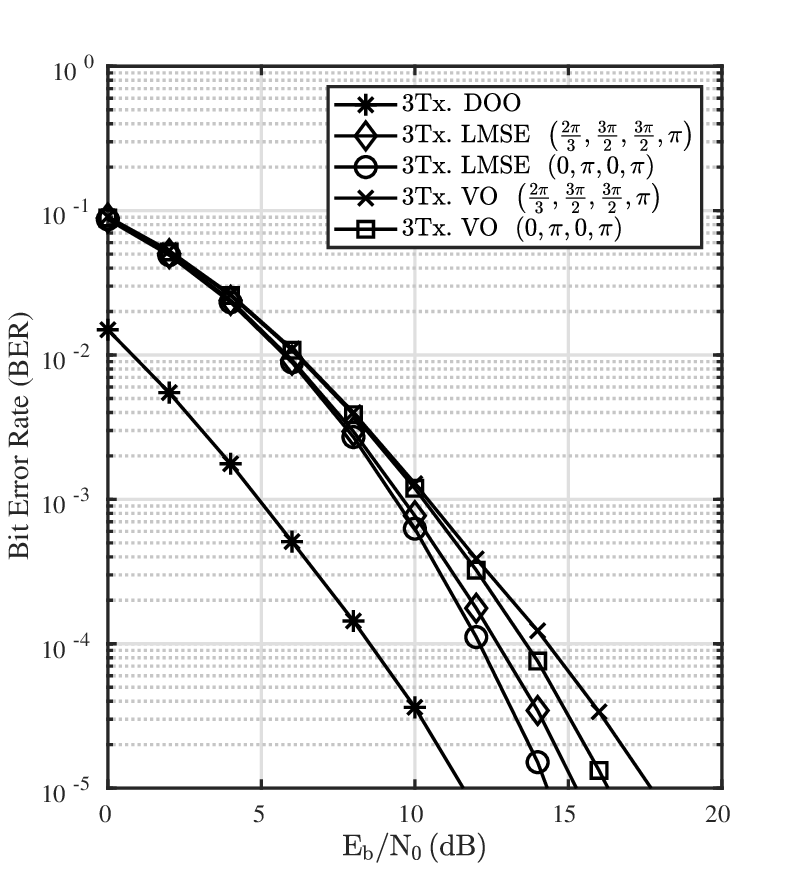}
		\caption{BER results for a $3$ user system with QPSK and three SIC Ordering Methods. System parameters: $K=3, N=32, M=4,d=[100,110,120]$m. The RIS phases are grouped in sets of four, with each set being $\left ( \frac{2\pi }{3} ,\frac{3\pi }{2}, \frac{3\pi }{2}, \pi \right ) $ or $\left ( 0 ,\pi, 0, \pi \right ) $}
\label{N=32}
\end{figure}

Next, we change the number of RIS elements $N$ to investigate its effects for these three SIC ordering methods. We start with $N=8$ and illustrate the results in Figure \ref{N=8} . The advantage of "DOO" over "LMSE" and "VO" is clear as shown in Figure \ref{N=8}.  Simulation results for $N=16$ are depicted in Figure \ref {N=16}, and  results for $N=32$  are depicted in Figure \ref{N=32}. From Figures \ref{N=8} -\ref{N=32} we see that for $N\ge 8$ SIC ordered by RIS beamforming optimization performs  the best in terms of  BER performance, while SIC ordered by LMSE performs slightly better than SIC ordered by channel variance.  The $E_b/N_0$ advantages of "DOO" over "LMSE $\left ( \frac{2\pi }{3} ,\frac{3\pi }{2}, \frac{3\pi }{2}, \pi, \dots, \frac{2\pi }{3} ,\frac{3\pi }{2}, \frac{3\pi }{2}, \pi   \right ) $" is around $1$dB at the BER of  $10^{-5}$ for $N=8$. The advantage increases to $1.8$dB for $N=16$, and $4.1$dB for N=32, indicating a more obvious advantage of Algorithm 3 over Algorithm 1 and Algorithm 2.


\subsubsection{Performance for Four Users System}
\begin{figure}[H] 
	\centering
	\includegraphics[scale=0.6]{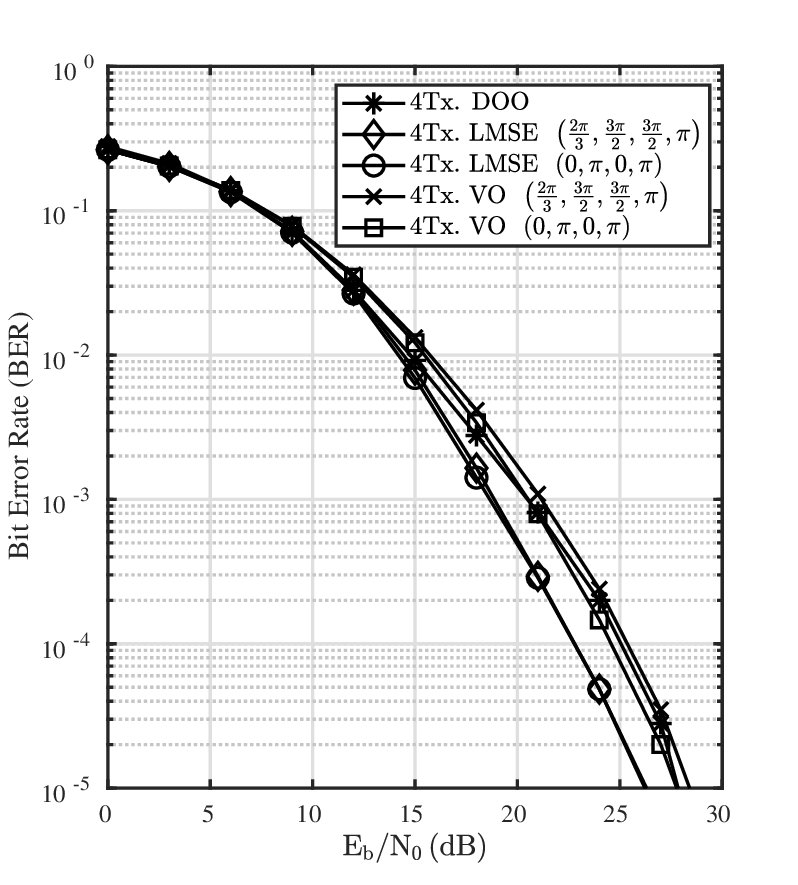}
	\caption{BER of a $4$ users system with QPSK,  for different SIC Ordering Methods. System parameters: $K=4, N=8, M=4,d=[100,110,120, 130]$m. The RIS phases are grouped in sets of four, with each set being $\left ( \frac{2\pi }{3} ,\frac{3\pi }{2}, \frac{3\pi }{2}, \pi \right ) $ or $\left ( 0 ,\pi, 0, \pi \right ) $}
	\label{4users_N=8}
\end{figure}

\begin{figure}[thbp!] 
		\centering
		\includegraphics[scale=0.6]{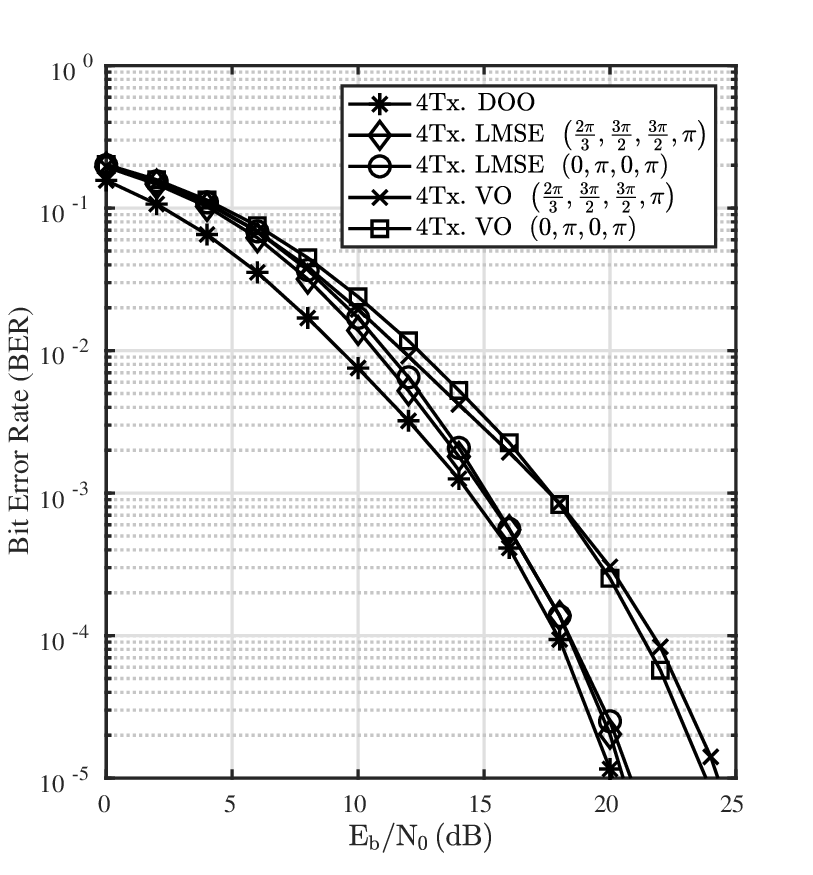}
		\caption{BER  of a $4$ users systems with QPSK, for different SIC Ordering Methods. System parameters: $K=4, N=16, M=4,d=[100,110,120, 130]$. The RIS phases are grouped in sets of four, with each set being $\left ( \frac{2\pi }{3} ,\frac{3\pi }{2}, \frac{3\pi }{2}, \pi \right ) $ or $\left ( 0 ,\pi, 0, \pi \right ) $}
\label{4users_N=16}
	\end{figure}

\begin{figure}[thbp!]
		\centering
		\includegraphics[scale=0.59]{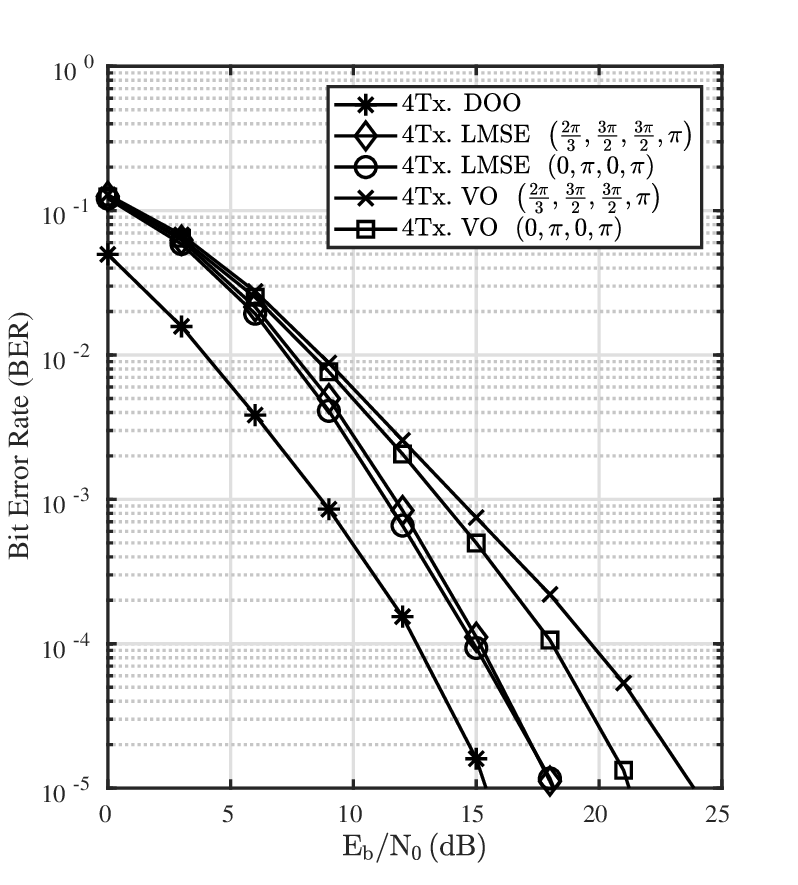}
		\caption{BER  of $4$ user systems with QPSK for  different SIC Ordering Methods. System parameters: $K=4, N=32, M=4,d=[100,110,120, 130]$m. The RIS phases are grouped in sets of four, with each set being $\left ( \frac{2\pi }{3} ,\frac{3\pi }{2}, \frac{3\pi }{2}, \pi \right ) $ or $\left ( 0 ,\pi, 0, \pi \right ) $}
\label{4users_N=32}
\end{figure}

We set the number of RIS elements $N$ to $8, 16, 32$ to simulate a system with four transmitting users.The simulation results are presented in Figures \ref{4users_N=8} -\ref{4users_N=32}. From Figure \ref{4users_N=8} we see that for $N=8$ the best BER performance is obtained by LMSE ordering. However, Figures \ref{4users_N=16} and \ref{4users_N=32} indicate that beamforming optimization provides the best BER performance. Hence also with four users and $N\geq 16 $ RIS beamforming optimization is superior.


 \subsubsection{Performance for Five Users System}
\begin{figure}[thbp!] 
	\centering
		\centering
		\includegraphics[scale=0.6]{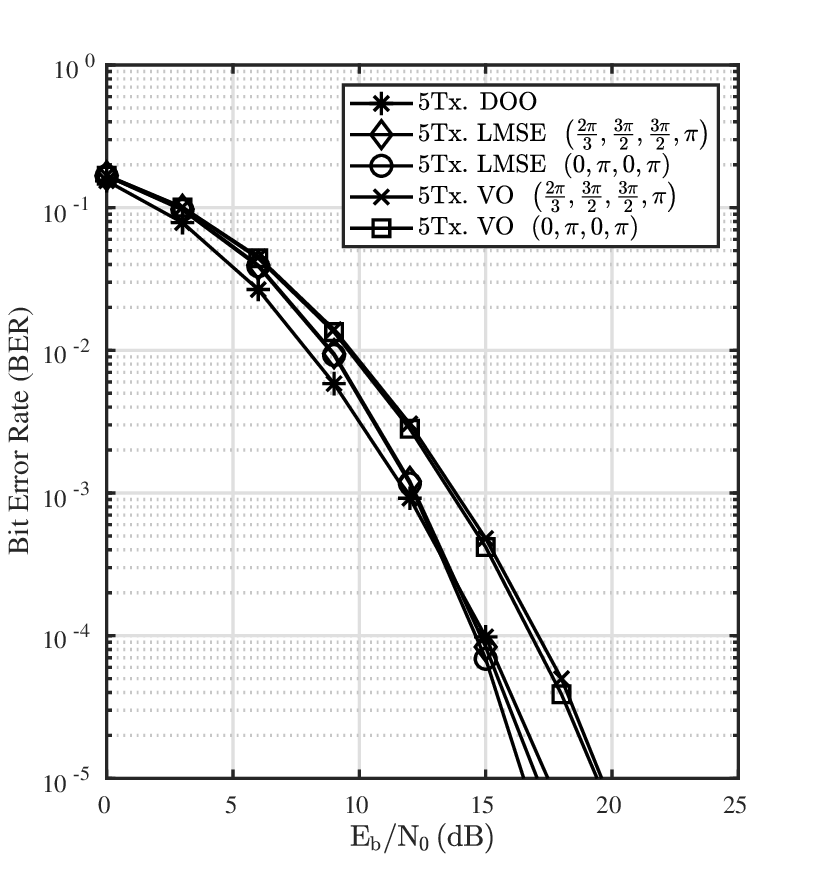}
		\caption{BER  of $5$ user systems with QPSK for three  SIC Ordering Methods. System parameters: $K=5, N=16, M=6,d=[100,110,120, 130, 140]$m. The RIS phases are grouped in sets of four, with each set being $\left ( \frac{2\pi }{3} ,\frac{3\pi }{2}, \frac{3\pi }{2}, \pi \right ) $ or $\left ( 0 ,\pi, 0, \pi \right ) $}
\label{5users_N=16}
	\end{figure}

\begin{figure}
		\centering
		\includegraphics[scale=0.6]{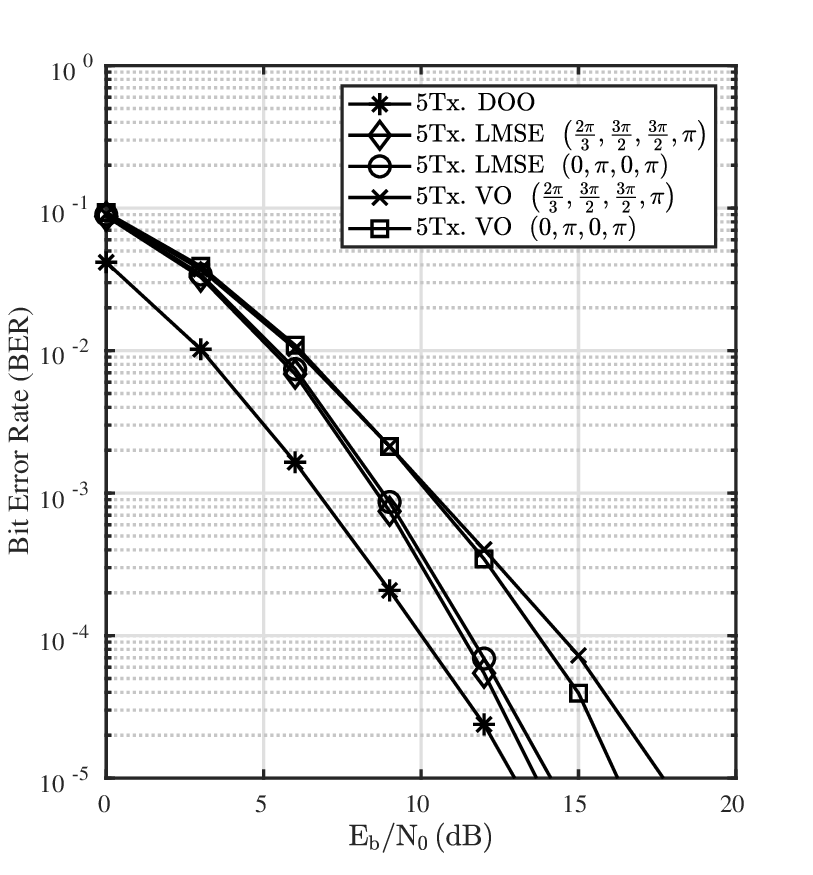}
		\caption{BER of $5$ user systems with QPSK for three  SIC Ordering Methods. System parameters: $K=5, N=32, M=6,d=[100,110,120, 130, 140]$m. The RIS phases are grouped in sets of four, with each set being $\left ( \frac{2\pi }{3} ,\frac{3\pi }{2}, \frac{3\pi }{2}, \pi \right ) $ or $\left ( 0 ,\pi, 0, \pi \right ) $}
\label{5users_N=32}
\end{figure}

\begin{figure}[thbp!] 
	\centering
	\includegraphics[scale=0.6]{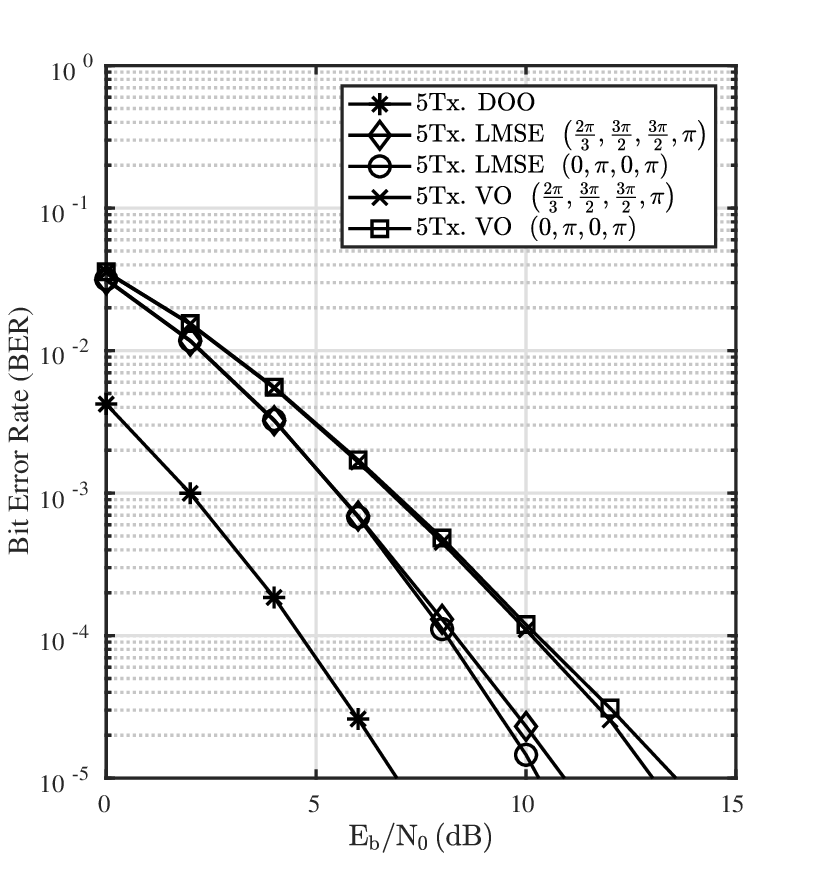}
	\caption{BER of $5$ user systems with QPSK for three SIC Ordering Methods. System parameters: $K=5, N=64, M=6,d=[100,110,120, 130, 140]$m. The RIS phases are grouped in sets of four, with each set being $\left ( \frac{2\pi }{3} ,\frac{3\pi }{2}, \frac{3\pi }{2}, \pi \right ) $ or $\left ( 0 ,\pi, 0, \pi \right ) $}
	\label{5users_N=64}
\end{figure}

We set the number of RIS elements to $16$, $32$, and $64$ to simulate a system with five users. The simulation results are depicted in Figures \ref{5users_N=16} - \ref{5users_N=64} . From these figures, we  see that  SIC ordered by RIS beamforming optimization (``DOO'' curves)  provides a notably better BER performance than SIC ordered by the other two methods. More specially, at a BER of $10^{-5}$, a $0.5$dB gain of "DOO"  compared with "LMSE $\left (\frac{2\pi }{3} ,\frac{3\pi }{2}, \frac{3\pi }{2}, \pi, \dots , \frac{2\pi }{3} ,\frac{3\pi }{2}, \frac{3\pi }{2}, \pi\right ) $"  is achieved for $N=32$, and a $3.5$dB gain of "DOO" compared with "LMSE $\left ( 0,\pi ,0,\pi, \dots, 0,\pi ,0,\pi  \right )$"  is achieved for $N=64$.

These results indicate  that RIS detection employing SIC ordered by beamforming optimization provides  a significant BER advantage when $N$ is sufficiently large ($N\geq16$) for various number of users. However for smaller values of $N$ LMSE ordering is better.


\section{SIC ordering  when  RIS reflection coefficients  suffer from magnitude variations}\label{mag_var}\label{mag_var}

In Section $\mathrm{\uppercase\expandafter{\romannumeral3}}$, we analyzed the performance of  SIC decoding ordering  assuming a constant RIS reflection magnitude is held constant  to $1$. However, practical reflection elements necessitate the consideration of energy loss due to hardware limitations\cite{abeywickrama2020intelligent}. Consequently, now we assume the amplitude of RIS reflecton coefficients are within the range $\left [0,1  \right ] $. Detailed descriptions of the revised optimization problem are provided. We then compare the BER performance of the three SIC methods for such reflection coefficients.

\subsection{Revised SIC Decoding Order Optimization for RIS reflection with amplitude variations}
Modifying  problem $\mathrm{(P3)}$,  we can derive the optimization algorithm which accounts for both amplitude and phase. After solving $\mathrm{(P3)}$ using Algorithm 3, we obtain a solution $\widehat{\bm\phi } =\begin{bmatrix}
	\hat{\phi }_1  & \hat{\phi }_2 & \dots  &\hat{\phi }_N
\end{bmatrix}^T$. However, according to \cite{abeywickrama2020intelligent}, there is a definite relation between the amplitude and phase for each RIS element, Thus we need to find the amplitudes and phases that are feasible for the  RIS model with reflection amplitude variations.  According to \cite{abeywickrama2020intelligent}, the phase-dependent amplitude of the $n$th diagonal element of the RIS reflective matrix $\mathbf{\Theta}=\mathrm{diag}(\eta _{1}e^{j\phi _{1}},\dots,\eta_{n}e^{j\phi _{n}},\dots,\eta _{N}e^{j\phi _{N}})$ can be expressed as:
\vspace*{-4mm}
\begin{equation}
	\eta_{n}   \left (\phi_{n}   \right ) =\left ( 1-\eta_{\mathrm{min} }  \right )\left ( \frac{\sin\left ( \phi_{n}- \phi   \right )  }{2}  \right ) ^{\alpha }+\eta_{\mathrm{min} } 
	\label{amplitude_phase}
\vspace*{-4mm}
\end{equation}
where $\eta_{\mathrm{min} }\ge 0 $ and it is the minimum RIS amplitude. Furthermore, $ \phi \ge 0 $  is the horizontal distance between $-\frac{\pi }{2} $ and $\eta_{\mathrm{min} }$. The exponent $\alpha \ge 0$ determines the steepness of the function. For $\eta_{\mathrm{min}}=1$, (\ref{amplitude_phase}) yields the ideal phase shift model with a constant amplitude of 1.  Similarly to the setup in \cite{abeywickrama2020intelligent}, we choose $\eta_{\mathrm{min} }=0.2$, $\alpha=1.6$, and $ \phi=0.43\pi$. Based on the optimization problem $\mathrm{(P3)}$ yielding $\widehat{\bm{\phi}}=\left [ \widehat{\phi }_1, \widehat{\phi }_2, \dots ,\widehat{\phi }_N  \right ] ^T$ we derive the optimal phase shift matrix with phases $\phi_{n}$ by solving problem (P4). 
\vspace*{-3mm}
\begin{equation}
	\begin{aligned} \label{P}
		\mathrm{(P4) :} &\textbf{for  n=1:N}\\
		& \min_{\phi _{n} } \left \| \eta \left ( \phi _{n}  \right ) e^{j\phi _{n}} - Ie^{j\hat{\phi} _{n}}\right \| ^{2}  \\
		&\begin{array}{l@{\quad}l@{}l@{\quad}l}
			s.t. & -\pi \le \phi _{n}  < \pi 			
		\end{array}
	\end{aligned}
\end{equation}
Under the assumption that the loss in RIS is small, we choose $I=1$.
Subsequently, we use the analytical expression (\ref{amplitude_phase}) to derive the actual amplitude $\eta_{n}$ based on $\phi_{n} $. The process is  summarized in algorithm 4.

\begin{algorithm}[H]
	%\textsl{}\setstretch{1.8}
	\renewcommand{\algorithmicrequire}{\textbf{Input:}}
	\renewcommand{\algorithmicensure}{\textbf{Output:}}
	\caption{ Optimization procedure for RIS with reflection amplitude variations based on Algorithm 3}
	\label{alg1}
	\begin{algorithmic}[1]
		\ENSURE  Decoding order $r_k(n)$
		\REQUIRE Optimized RIS phase $\widehat{\bm{\phi}}=\left [ \widehat{\phi }_1, \widehat{\phi }_2, \dots ,\widehat{\phi }_N  \right ] ^T$ after solving problem (P3)
		\STATE Initialization:$\left\{ \eta_{\mathrm{min}}=0.2,\alpha=1.6, \phi=0.43\pi \right\}$
		\STATE Calculate phase vector $\bm{\phi}=\left[ \phi_1, ...\phi_N \right]$ by solving problem (P4)
		\STATE  Calculate RIS amplitude vector $\mathbf{\eta}$ by expression (\ref{amplitude_phase})
		\STATE  Derive RIS diagonal matrix $\mathbf{\Theta }$
		\STATE Calculate all combined channel gains $\left \{ \left \| \mathbf{H}_{kM}^T\mathbf{\Theta }\mathbf{H}_K\mathbf{P} \right \|_F ^{2},k\in \left \{ 1,2,\dots K \right \}  \right \}$ and rank them in descending order for each channel
	\end{algorithmic}  
\end{algorithm}
\enlargethispage*{20mm}
We  employ the fmincon function of MATLAB for constraint minimization implied by problem P4. The default method (Internal-point method\cite{nocedal1999numerical}) is used to solve the non-linear optimization.

Next we present  simulations results to illustrate the performance of such schemes.  For each $E_b/N_0$ the BER is estimated  based on at least 500-bit errors. The transmitted bits are generated randomly, and  BER results are averaged over 1000 channel realizations.

\subsection{Performance results for RIS with amplitude variations in reflection coefficients}

Next we present computer simulation results for SIC with three ordering methods when  the RIS reflection coefficients are non-ideal and suffer from amplitudee variations.

\subsubsection{Performance for Three User System}

\begin{figure}[H] 
	\centering
	\includegraphics[scale=0.6]{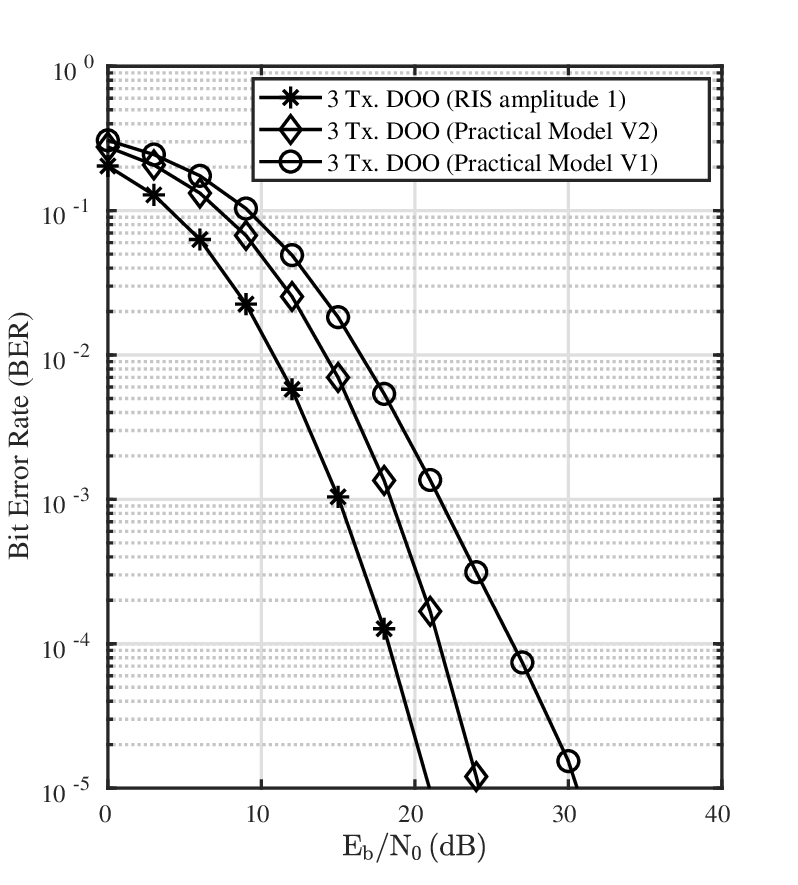}
	\caption{BER  of $3$ user systems with QPSK  for optimization of  SIC ordering based on Algorithm 3, and 4.  System parameters: $K=3, N=8, M=4,d=[100,110,120, 130, 140]$m. The RIS phases are grouped in sets of four, with each set being $\left ( \frac{2\pi }{3} ,\frac{3\pi }{2}, \frac{3\pi }{2}, \pi \right ) $ or $\left ( 0 ,\pi, 0, \pi \right ) $}
	\label{var_ampl1}
\end{figure}

\begin{figure}[thbp!] 
		\centering
		\includegraphics[scale=0.6]{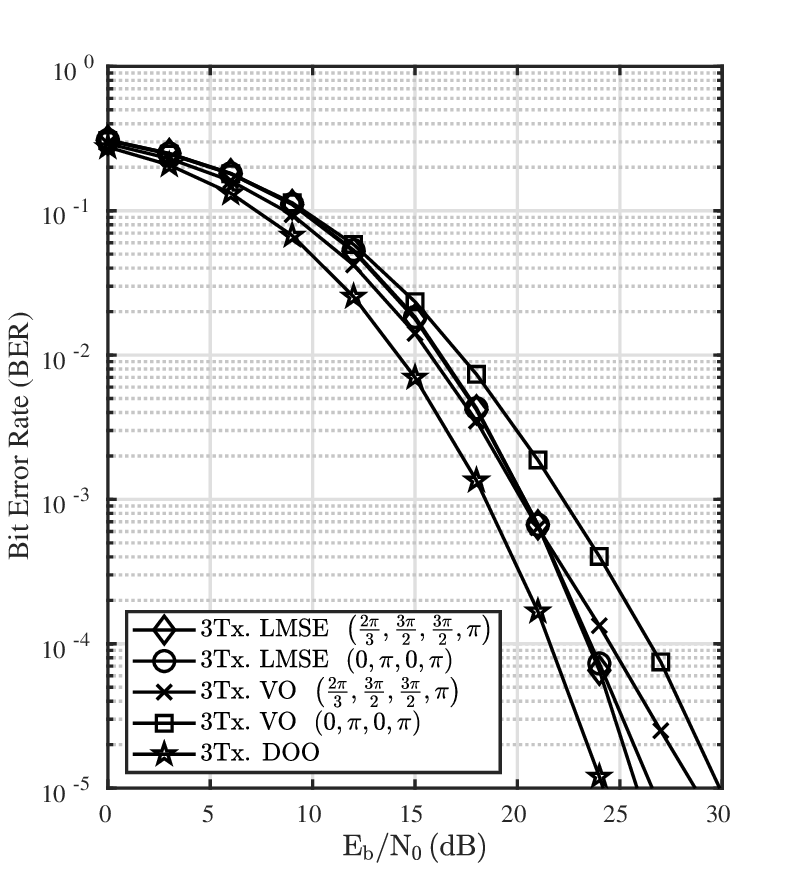}
		\caption{ BER of $3$ user systems with QPSK for three SIC ordering methods. System parameters: $K=3, N=8, M=4, d=[100,110,120]$m. The RIS phases are grouped in sets of four, with each set being $\left ( \frac{2\pi }{3} ,\frac{3\pi }{2}, \frac{3\pi }{2}, \pi \right ) $ or $\left ( 0 ,\pi, 0, \pi \right ) $}
		\label{var_ampl2}
	\end{figure}

\begin{figure}[thbp!]
		\centering
		\includegraphics[scale=0.6]{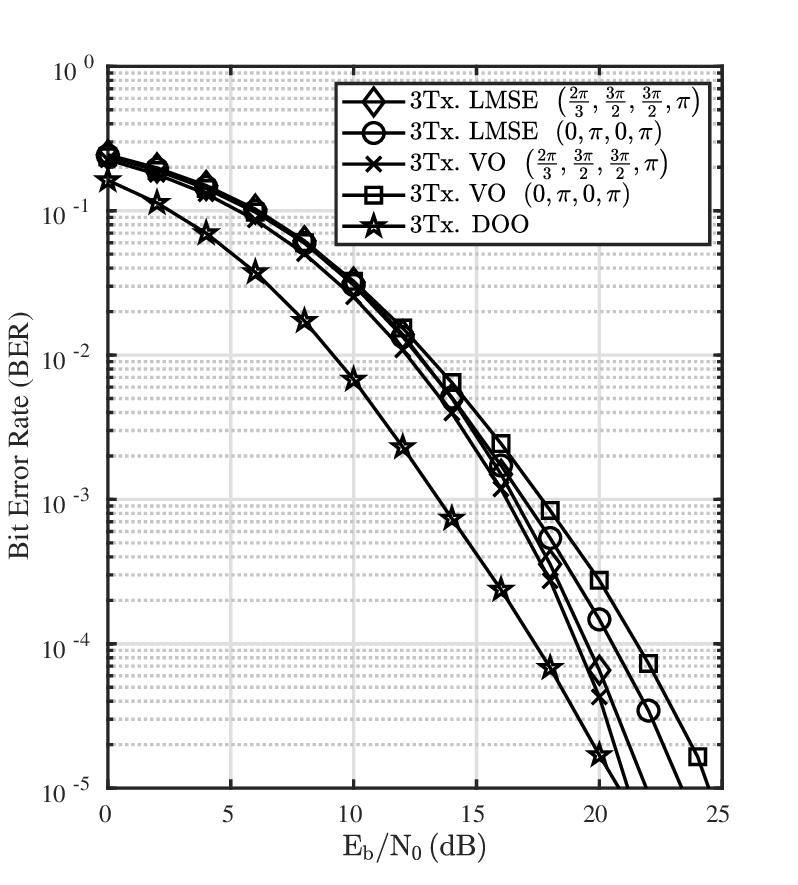}
		\caption{BER  of $3$ user systems with QPSK for three  SIC ordering methods. System parameters: $K=3, N=16, M=4, d=[100,110,120]$m. The RIS phases are grouped in sets of four, with each set being $\left ( \frac{2\pi }{3} ,\frac{3\pi }{2}, \frac{3\pi }{2}, \pi \right ) $ or $\left ( 0 ,\pi, 0, \pi \right ) $}
\label{var_ampl3}
\end{figure}

\begin{figure}[H] 
	\centering
	\includegraphics[scale=0.6]{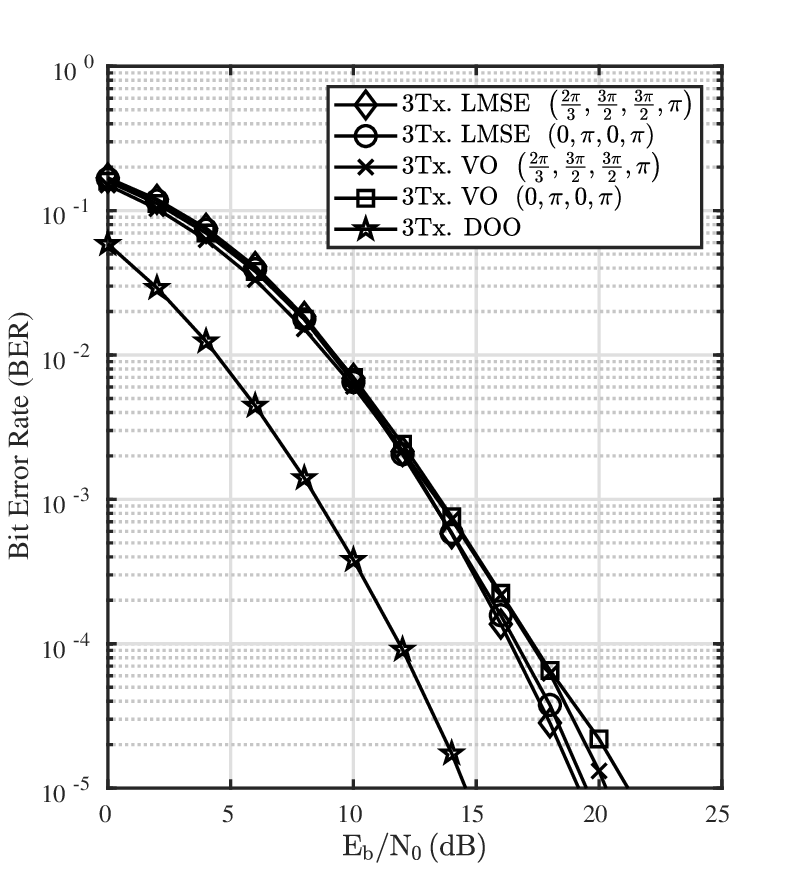}
	\caption{BER  of $3$ user systems with QPSK for three SIC ordering methods. System parameters: $K=3, N=32, M=4, d=[100,110,120]$m. The RIS phases are grouped in sets of four, with each set being $\left ( \frac{2\pi }{3} ,\frac{3\pi }{2}, \frac{3\pi }{2}, \pi \right ) $ or $\left ( 0 ,\pi, 0, \pi \right ) $}
	\label{var_ampl4}
\end{figure}

We consider a system with three users with $N$, the numbers of RIS reflecting elements, set to $8$, $16$ and $32$. 
\enlargethispage*{40mm}
We start with Algorithms 3 and 4 by analyzing the  three cases: $(1)$ Algorithm 3 with  fixed RIS coefficients amplitude set to 1; $(2)$ Algorithm 4 with varying RIS amplitudes; and $(3)$ Algorithm 3 with a varying RIS amplitudes. In case $(3)$, we use the same RIS phases as in case $(1)$ to  calculate the RIS amplitude by  (\ref{amplitude_phase}), and then apply the calculated amplitudes to the system. We label the curve for each case as "3 Tx. DOO (RIS ampl. 1)", "3 Tx. DOO + Alg.4 (RIS var. ampl.)", and "3 Tx. DOO (RIS var. ampl.)", respectively. These  three curves are depicted in Figure \ref{var_ampl1}. We see that  case (1) achieves the best BER performance. However, when the RIS amplitudes vary case (3) achieves the worst performance. When the RIS ampitudes vary, case (2) shows a performance advantge of algorithm 4 over algorithm 3, by 2.7 dB at a BER of $10^{-4}$.
\vspace*{-7mm}
\newpage
For  $N=8$, $N=16$ and $N=32$, we present results for Algorithms 1,2 and 4 in Figures \ref{var_ampl2} - \ref{var_ampl4} for RIS with varying amplitudes. 
Comparing the 'DOO' curve between Figures \ref{N=16} and \ref{var_ampl3} , and between Figures \ref{N=32} and \ref{var_ampl4}, shows  that RIS amplitude-based SIC order optimization performs worse than RIS fix amplitude 1-based SIC optimization, with losses of $5$ dB at a BER of $10^{-4}$ and $6.5$ dB at a BER of $10^{-5}$ for N=16. Additionally, there is a loss of  $4.5$ dB at a BER of $10^{-4}$ and $3.3$ db at $10^{-5}$ for N=32.

Figures  \ref{var_ampl2} - \ref{var_ampl4}  show  the SIC ordered by RIS optimization results in the best BER performance, while SIC ordered by LMSE performs approximately the same as SIC ordered by channel variance. At a BER of $10^{-4}$, the performance gap between "DOO" and "LMSE" is about $1.8$ dB for N=8 , $1.5$ dB for N=16, and  $5$ dB for N=32.

Comparing the BER results for different number of RIS reflecting elements $N$ , shows that  the configuration with $N=32$ achieves the best BER performance, whereas the configuration with $N=8$ exhibits the worst performance.  At a BER of $10^{-4}$, the performance gaps between $N=32$ with respect to  $N=16$ and  $N=8$ are $5.2$ dB, $9.6$ dB, respectively. At a BER of $10^{-5}$, these gaps increase to $6.4$ dB, $9.7$ dB, respectively.  

To summarize,  RIS with amplitude variations leads to performance losses compared to the ideal case of fixed amplitudes. These losses are reduced when using algorithm 4 that acoounts for such amplitude variations in the optimization process. The BER performance consistently improves as $N$ increases.

\vspace*{-5mm}

\vspace*{-20mm}

\vspace*{-3mm}
\subsubsection{Performance for Four User System}

\begin{figure}[thbp!] 
	\centering
		\includegraphics[scale=0.6]{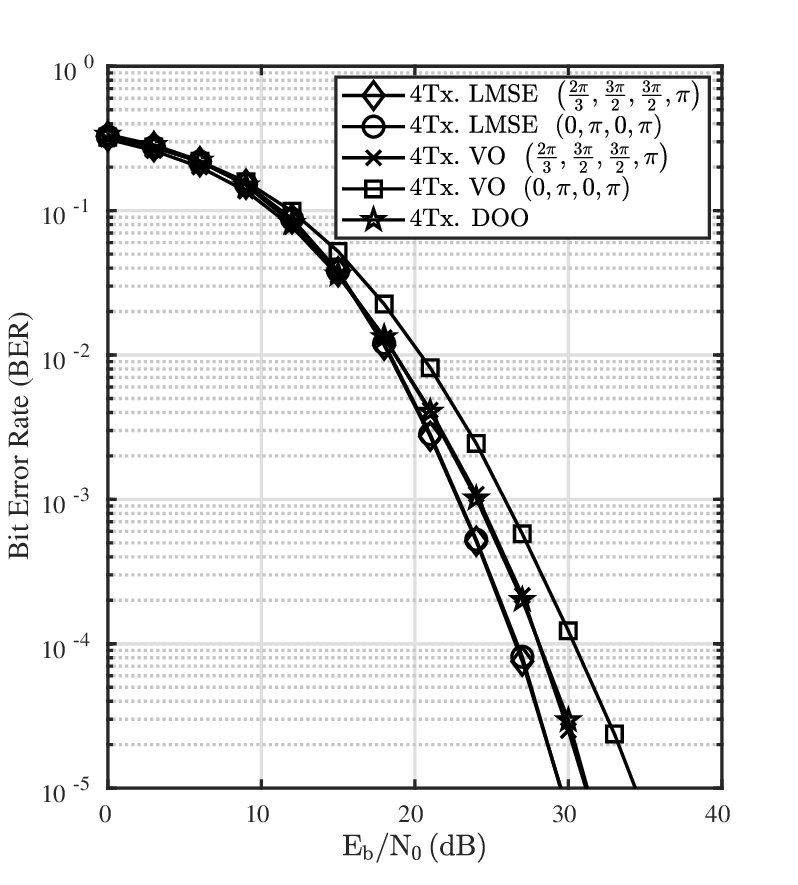}
		\caption{BER  of $4$ user systems with QPSK, for three  SIC ordering ethods. System parameters: $K=4, N=8, M=4, d=[100,110,120,130]$m. The RIS phases are grouped in sets of four, with each set being $\left ( \frac{2\pi }{3} ,\frac{3\pi }{2}, \frac{3\pi }{2}, \pi \right ) $ or $\left ( 0 ,\pi, 0, \pi \right ) $}
		\label{ampl_var2_4users_N=8}
	\end{figure}

\begin{figure}[thbp!]
		\centering
		\includegraphics[scale=0.6]{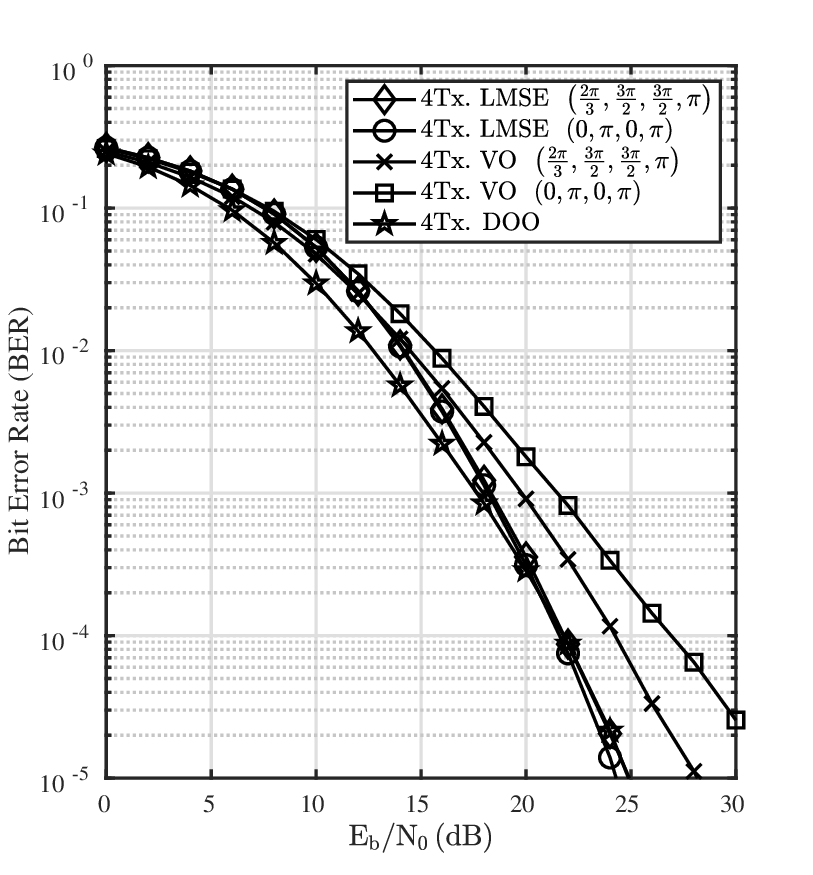}
		\caption{BER of $4$ user systems with QPSK for three  SIC ordering methods. System parameters: $K=4, N=16, M=4, d=[100,110,120,130]$m. The RIS phases are grouped in sets of four, with each set being $\left ( \frac{2\pi }{3} ,\frac{3\pi }{2}, \frac{3\pi }{2}, \pi \right ) $ or $\left ( 0 ,\pi, 0, \pi \right ) $}
	\label{ampl_var2_4users_N=16}
\end{figure}

\begin{figure}[H] 
	\centering
	\includegraphics[scale=0.6]{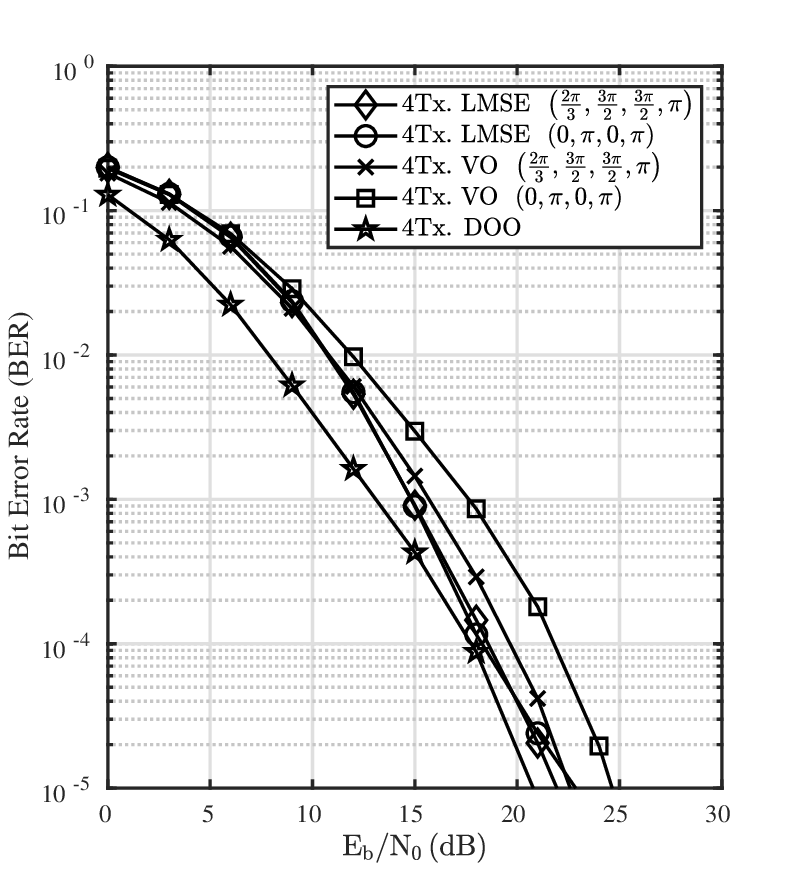}
	\caption{BER  of $4$ user systems with QPSK for  three  SIC ordering methods. System parameters: $K=4, N=32, M=4, d=[100,110,120,130]$m. The RIS phases are grouped in sets of four, with each set being $\left ( \frac{2\pi }{3} ,\frac{3\pi }{2}, \frac{3\pi }{2}, \pi \right ) $ or $\left ( 0 ,\pi, 0, \pi \right ) $}
	\label{ampl_var2_4users_N=32}
\end{figure}

We consider now the  three SIC ordering algorithms in a system with four users, and set the number of RIS reflecting elements $N$ to $8$, $16$ and $32$. 
We assume RIS reflection coeffinents with amplitude variations, and present the results in  Figures \ref{ampl_var2_4users_N=8} -  \ref{ampl_var2_4users_N=32} for algorithms 
1,2,4.

Comparison of the 'DOO' curve between Figure \ref{4users_N=8} and Figure \ref{ampl_var2_4users_N=8}, Figure \ref{4users_N=16} and Figure \ref{ampl_var2_4users_N=16},  Figure \ref{4users_N=32} and Figure \ref{ampl_var2_4users_N=32}, shows that  SIC ordered by RIS optimization following algorithm 4 when RIS elements suffer from amplitude variations performs worse than  VO, LMSE and phase optimization (algorithm 3) over constant amplitude  RIS elements channels. We observe losses of around $3$dB at BER of $10^{-4}$ for $N=8$. We also see losses of around $4$dB at the BER of $10^{-4}$  for $N=16$, and around $5$dB at the BER of $10^{-4}$  for $N=32$.
Furthermore Figures \ref{ampl_var2_4users_N=8} -  \ref{ampl_var2_4users_N=32} reveal that with RIS coefficient amplitude variations, for $N=8$ VO and LMS ordering perfom better than the optimization method of algorithm 4 (DOO). However for $N\ge 16$ the optimization method of algorithm 4 provides the best performance.

In summary, amplitude variations in RIS elements results in performance losses compared to ideal amplitude cases. For $N<16$, SIC ordered by the  LMSE method attains the best BER performance. However for $N\ge16$  SIC ordered by the optimization method of algorithm 4 achieves the best BER performance. To avoid repetitions,  we do not include results for 5 or more users, since they follow a similar trend as the results for 4 users

	%\vspace*{-10mm}

\enlargethispage*{20mm}

\section{Conclusions}\label{concl}

This paper considers an RIS aided multi-way communication system with $K$ user terminals and $N$ RIS elements. Each user is equipped with one transmit antenna and $M$ receive antennas. For data detection  we employ Successive Interference Cancellation (SIC) with three decoding order approaches: 1) channel coefficient variance ordering (Algorithm 1), 2) LMSE ordering (Algorithm 2), and 3) RIS beamforming optimization (Algorithm 3). We assume a system of 4 receive antennas per user, with 1 - 5 users , and initially study the BER behavior  of Algorithms 1 and 2. Numerical results show that the achievable diversity orders are  close to the asymptotic diversities, with the achievable diversities of SIC ordered by LMSE being slightly larger than that of SIC ordered by channel variances.
It is shown that  SIC ordered by LMSE achieves the best BER  results with fixed RIS reflection coefficients amplitudes and phases.

For Algorithm 3 we  assume three scenarios: 3 transmitting users, 4 transmitting users and 5 transmitting users.. Within each scenario, we vary the number of RIS elements and compare the numerical results of Algorithms $1-3$. Initially the amplitudes for each RIS element is set to 1. As the number of RIS elements increases from 4 to 64, it is observed that Algorithm 2 outperforms the others when $N$ is small; however, Algorithm 3 significantly outperforms the others when $N \ge 16$ . Furthermore, the BER decreases as the number of RIS elements increases.

Finally, we derive a beamforming optimization algorithm for SIC ordering  when the RIS reflecting elements suffer from varying amplitudes.. This technique allows us to account for the  behavior of RIS elements  in practical deployment scenarios. Simulation  results indicate performance gains over ordering methods that do not account for RIS amplitude variations in reflection coefficients.  The advantages of RIS beamforming optimizations hold true for a large number of RIS elements even when there are variations in RIS reflection amplitudes.

Future work could extend this research by exploring SIC methods for multi-way RIS based  relaying with other modulation schemes, such as high order QAM. Additionally, the optimization problem can be improved to achieve optimal BER performance regardless of the number of RIS elements. Extending such techniques to adapt to varying channel conditions could be a useful further extension of our work

\end{spacing}

 \bibliographystyle{ieeetr}
 \bibliography{IEEEabrv,references}

\end{document}